\documentclass[aps,prb,reprint,twocolumn,showpacs,superscriptaddress]{revtex4-2}
\usepackage{physics}
\usepackage{amsmath}
\usepackage{graphicx}
\usepackage{lmodern}
\usepackage{amsmath}
\usepackage{color}
\usepackage{amssymb}
\usepackage{bm}
\usepackage{braket}
\usepackage{mathtools}
\usepackage{afterpage}
\usepackage{dsfont}

\newcommand{\up}{\uparrow}
\newcommand{\down}{\downarrow}
\newcommand{\br}{\bm{r}}
\newcommand{\bk}{\bm{k}}
\newcommand{\bp}{\bm{p}}
\newcommand{\bq}{\bm{q}}

\newcommand{\bdelta}{\bm{\delta}}
\newcommand{\bphi}{\bm{\phi}}

\usepackage[compact]{titlesec}         
\titlespacing{\section}{5pt}{5pt}{5pt} 
\AtBeginDocument{
  \setlength\abovedisplayskip{5pt}
  \setlength\belowdisplayskip{5pt}}

\newcommand{\beginsupplement}{%
   \setcounter{equation}{0}
   \pagebreak
\onecolumngrid
\setcounter{figure}{0}
\setcounter{table}{0}
\setcounter{page}{1}%
\renewcommand{\thefigure}{S\arabic{figure}}
}
\newcommand{\be}{\begin{equation}}
\newcommand{\ee}{\end{equation}}

\usepackage[dvipsnames]{xcolor}
\usepackage[colorlinks=true]{hyperref}
\hypersetup{
    colorlinks=true,
    linkcolor=blue,
    citecolor=blue,
    urlcolor=blue
} 

\makeatletter
\def\maketitle{
\@author@finish
\title@column\titleblock@produce
\suppressfloats[t]}
\makeatother

\begin{document}
\title{Stability of Quasiperiodic Superconductors}
\author{Nicole S. Ticea}
\affiliation{Department of Applied Physics, Stanford University, Stanford, CA 94305, USA}
\author{Julian May-Mann}
\affiliation{Department of Physics, Stanford University, Stanford, CA 94305, USA}
\author{Jiewen Xiao}
\affiliation{Department of Condensed Matter Physics, Weizmann Institute of Science, Rehovot 76100, Israel}
\author{Erez Berg}
\affiliation{Department of Condensed Matter Physics, Weizmann Institute of Science, Rehovot 76100, Israel}
\author{Trithep Devakul}
\affiliation{Department of Physics, Stanford University, Stanford, CA 94305, USA}
\date{\today}

\begin{abstract}
We study the effects of quasiperiodicity on the stability of conventional and unconventional superconductors. Quasiperiodicity is modelled using the three-dimensional Aubry-André (AA) model, a system in which electrons are coupled to a translation-symmetry-breaking potential that is incommensurate with the underlying lattice. Upon increasing the strength of the quasiperiodic potential, the single-particle eigenstates undergo a transition from a ballistic to diffusive character. Here, we study the instability of the model towards superconductivity. We find that in the ballistic regime, the system is unstable towards both $s$-wave and $p$-wave superconductivity. In contrast, only the conventional $s$-wave instability survives in the intermediate diffusive regime. Our result suggest a version of Anderson’s theorem for quasiperiodic systems, relating the normal state dynamics to the stability of conventional and unconventional superconductivity. These findings are relevant vis-à-vis recent studies of superconductivity in quasiperiodic moiré structures. 
\end{abstract}
\maketitle
\section{INTRODUCTION} 
\label{sec:introduction}
Quasiperiodic materials have been studied by crystallographers since the early 1980s \cite{mackay,mackay2}, though evidence of a natural quasicrystal did not arise until 2009 with the discovery of icosahedrite \cite{natural_qc}. Since then, interest in the field has surged with the realization of quasicrystals in a wide range of tunable systems, including optical \cite{optical1,optical2,optical3,optical4,optical5,optical6,optical7} and photonic lattices \cite{photonic1,photonic2,photonic3,photonic4,photonic5}, cavity polaron devices \cite{polaron}, and moiré materials \cite{ huang2019moire, moire1}. With tunability comes the opportunity to study how quasiperiodicity not only disrupts or stabilizes known physics, but can also engender new phenomena \cite{qhi, topo, chen2020higher, else2021quantum, jagannathan2021fibonacci}.  

In this paper, we seek to understand how quasiperiodicity affects conventional ($s$-wave) and unconventional (non-$s$-wave) superconductivity. 
We are motivated, in part, by the existence of superconductivity in moiré systems \cite{cao2018, balents2020,yankowitz2019,lu2019,chen2019}. Although all twisted moiré graphene systems exhibit a quasiperiodic arrangement of atoms at generic twist angles, the low-energy electronic behaviour in most paradigmatic structures (e.g., twisted bilayer graphene) is determined by an emergent long-wavelength moiré periodicity \cite{bistritzer_macdonald}. 
Recent studies, however, suggest that it is possible to realize superconductivity in an incommensurate trilayer system, in which even the low energy physics is quasiperiodic \cite{aviram, yang2023multi}. 
This discovery may yield new insight into the nature of superconductivity in twisted trilayer graphene specifically, and in multilayer graphene devices more generally. Namely, if the superconductivity found in quasiperiodic twisted trilayer graphene is the same as that found in other twisted trilayer graphene systems \cite{park2021tunable, hao2021electric, cao2021pauli, liu2022isospin,  kim2022evidence, talantsev2022compliance, park2022}, this would suggest an intrinsic stability of the superconducting state to quasiperiodicity. 
It is therefore of crucial importance to identify any fundamental differences in the stability of conventional and unconventional superconductivity to quasi-periodicity.

\begin{figure}[h]
\centering
\includegraphics[width=0.48\textwidth]{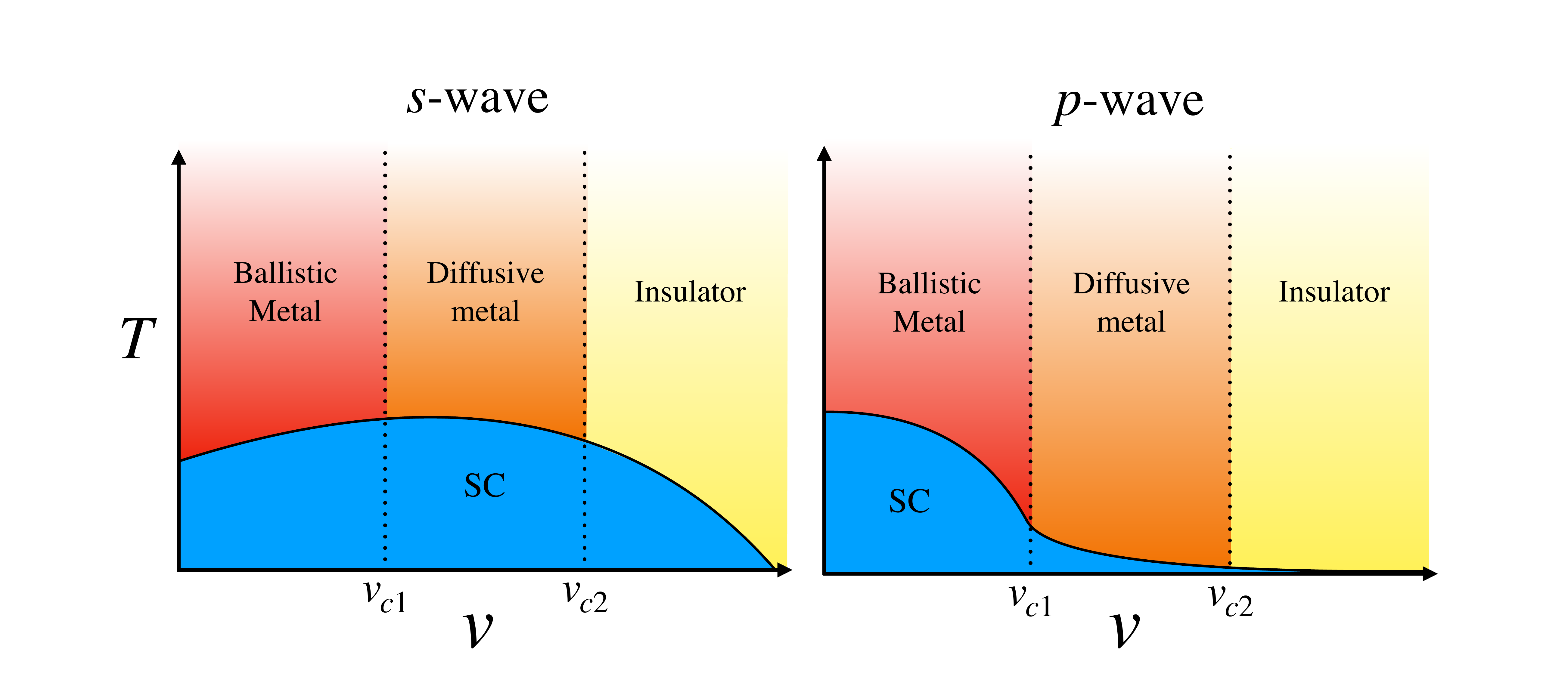}
\caption{Schematic phase diagram for $s$- and $p$-wave superconductivity in the $3$D Aubry-André model as a function of quasiperiodic potential strengh, $v$. In the normal state, a ballistic-to-diffusive and diffusive-to-localized transition occurs at $v = v_{c1}$ and $v = v_{c2}$ respectively.}
\label{fig:phase_diagram}
\end{figure} 
Because translational symmetry is broken in systems with either quasiperiodicity or random disorder, one might expect quasiperiodic systems to behave similarly to those with disorder. Quasiperiodic lattices, however, possess two or more spatially periodic orders that are incommensurate with each other \cite{levine1986quasicrystals}. Because of these periodic orders, quasiperiodic systems---unlike disordered systems---are highly structured in momentum space; the diffraction pattern of the ideal, infinite quasicrystal can be characterized by a self similar arrangement of Bragg peaks, densely tiled throughout reciprocal space \cite{levine1984}. A paradigmatic quasiperiodic system is the Aubry-André (AA) model in one spatial dimension (1D) \cite{aubry1980analyticity}, a model of fermions on a lattice subject to a periodically-modulating onsite potential that is incommensurate with the lattice. As the potential strength is tuned, this system exhibits a transition at finite potential strength from an phase with ballistic states (i.e., momentum-space localized states) to a phase with real-space localized states. In contrast, randomly disordered systems in one dimension tend towards localization for infinitesimal disorder. The AA model can also be generalized to 3D \cite{trithep}, where it displays two finite-potential-strength transitions: first from a ballistic to a diffusive phase, and then from the diffusive phase to a real-space-localized phase. For random systems in 3D, the ballistic-to-diffusive transition occurs at infinitesimal disorder strength. 

A number of works have considered conventional and unconventional superconductivity in other quasiperiodic systems, such as the Penrose lattice \cite{sakai2017, araujo2019conventional, sakai2019exotic, cao2020kohn, takemori2021, ghadimi2021topological, nagai2022intrinsic, takemori2024superconducting}. Conventional superconductivity has also been analyzed in the 1D AA model, where an enhancement of superconductivity is observed near the ballistic-to-localized transition 
\cite{enhanced_sc_1d}\footnote{This enhancement near criticality can be understood as a trade-off between maximizing the pairing strength (which is largest when the electrons are locally confined) and global phase coherence (which is low in the localized phase)}. An advantage of working with the AA model is that the quasiperiodicity can be tuned by adjusting the strength of the potential, in the same way that one can study the effects of random disorder simply by tuning the strength of the disordered potential. This makes it possible to directly compare the effects of disorder and quasiperiodicity on superconductivity. In randomly disordered systems, low levels of non-magnetic disorder do not affect $s$-wave superconductivity. This is Anderson's theorem for disordered superconductors \cite{Anderson}. However, unconventional superconductors, such as $p$- and $d$-wave superconductors, are fragile to even small amounts of random disorder in three-dimensions \cite{sigrist1991phenomenological, keles2014}. This motivates us to ask if quasiperiodicity affects conventional and unconventional superconductivity in the same manner as does random disorder. 

In the three-dimensional (3D) AA model, we find that weak-coupling instabilities of both conventional \text{and} unconventional superconductors are robust to small, but finite, quasiperiodicity. For unconventional superconductivity, the instability vanishes once the strength of the quasiperiodic potential exceeds a critical value. This critical value coincides with the transition from the ballistic to the diffusive phase of the normal state; that is to say, weak coupling unconventional superconductivity only occurs in the ballistic phase of the 3D AA model. For conventional superconductivity, the weak coupling instability persists throughout the ballistic and diffusive phases. For finite strength couplings, we find that the $T_c$ for unconventional superconductivity is strongly suppressed in the diffusive and localized regimes, while the $T_c$ for conventional superconductivity only starts to become suppressed in the localized phase. This is illustrated in Fig.~\ref{fig:phase_diagram}. For both conventional and unconventional superconductivity, we find that the suppression of $T_c$ is primarily due to phase fluctuations. Taken together, these results suggest a generalization of Anderson's theorem of disordered superconductors to quasiperiodic systems. 

\section{MODEL}
\label{sec:method}
Our starting point is the 3D generalization of the AA model, first introduced in Ref.~\cite{trithep}:
\begin{align}
    H = \sum_{\sigma}\sum_{\br}\sum_{i=1}^3(e^{i\phi_i}c^\dagger_{\br+\hat{u}_i\sigma}c_{\br\sigma}+ h.c.) + H_\text{QP}\label{hamiltonian}
\end{align}
The sum is over all sites $\br$ on an $L\times L\times L$ cubic lattice, $\{\phi_i\}$ are arbitrary phases that twist the periodic boundary conditions in all three directions, $\{\hat{u}_i\}$ are the lattice basic vectors, and $c_{\br\sigma}$ annihilates a fermion on site $\br$ with spin $\sigma = \uparrow,\downarrow$. 

The quasiperiodic potential is specified by
\begin{equation}
    H_\text{QP} = 2v\sum_{\br} \sum_{i=1}^3\cos\left(2\pi\sum_{j=1}^3 B_{ij}r_j + \phi_i \right) n_{\br}, \label{HQP}
\end{equation}
where $n_{\br} = \sum_{\sigma}c^\dagger_{\br\sigma}c_{\br\sigma}$, and $v$ controls the strength of the quasiperiodic potential. Here, the phases $\{\phi_i\}$ amount to an overall arbitrary shift of the cosine potential. These phases are taken to be the same in Eq.~\ref{hamiltonian} and ~\ref{HQP} so that this model is self-dual upon transforming from real to momentum space and sending $v\to 1/v$ \cite{trithep}. In principle, however, the phases appearing in the hopping and potential terms can be independent. 

The matrix $\mathbf{B}$ determines the characteristics of this model most relevant to this paper. If we wish to define a quasiperiodic system, then we impose that $\sum_j B_{ij} r_j \notin \mathbb{Z}^3$ for all $\br \neq 0$. Here we take $\mathbf{B}= Q\mathbf{R}(\theta)$, where $Q=\frac{\sqrt{5}-1}{2}$ is the golden ratio and $\mathbf{R}(\theta)$ can be thought of as a reflection about the line $y=z$ followed by three Euler rotations ($Y_\theta X_\theta Z_\theta$) around an angle $\theta$, which we take here to be $\pi/7$ \footnote{Specifically, letting $c\equiv\cos\theta$ and $s\equiv\sin\theta$, we define $\mathbf{R}(\theta) = \begin{pmatrix}
c^2+s^3 & cs & cs^2-cs\\
cs & -s & c^2\\
cs^2-cs & c^2 & c^2s+s^2
\end{pmatrix}$}. In practice, it is useful to define the AA model on a finite-size system on periodic boundary conditions. To do so, we must enforce periodicity in system size $L$, which amounts to picking a rational approximant for $\mathbf{B}$ such that $L\mathbf{B}$ is an integer-valued matrix. We also require the approximate form of $\mathbf{B}$ to satisfy $\gcd (\det (L \mathbf{B}), L) = 1$; as a result, the only translation symmetry preserved by the cosine potential is $\br \rightarrow \br + L \bm{n}$, with $\bm{n}\in \mathbb{Z}^3$. On top of breaking translation symmetry, the quasiperiodic potential also breaks rotation symmetry around any axis. The model, however, remains invariant under inversion, $\br\to -\br$ when the phases $ \{ \phi_i \}$ are equal to $0$ or $\pi$. 

\begin{figure}[h]
\centering
\includegraphics[width=0.48\textwidth]{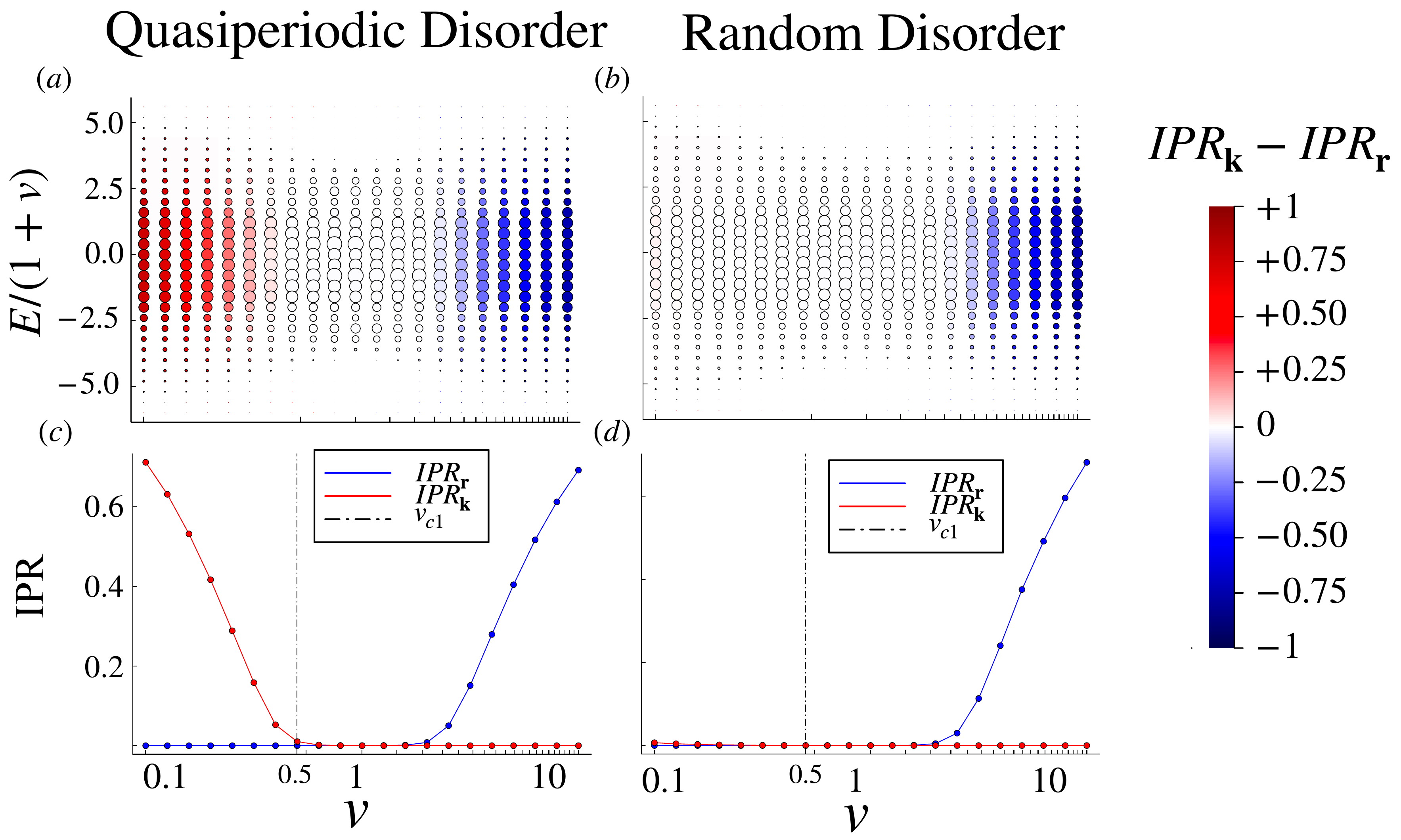}
\caption{Top row: The difference between the real- and momentum-space IPRs for the eigenstates of the Hamiltonian described in Eq.~\ref{hamiltonian}. At low $v$, the state is initially localized in momentum space (red); as $v$ is increased, the state becomes real-space localized (blue). The size of the data points corresponds to the density of states. Bottom row: the momentum (red) and real space (blue) IPRs as a function $v$ along the $E/(1+v) = 0.75$ linecut.}
\label{fig:IPR}
\end{figure}

The real-to-momentum space duality discussed earlier possesses a self dual point at $v_\text{SD}=1$. Previous studies \cite{trithep} on this model report an extended diffusive phase, which is not localized in either real or momentum space. It is bracketed on either side by a finite ballistic phase ($v<v_{c_1}$) and a (real-space) localized phase ($v>v_{c_2}$). The mean-square displacement of an initially localized particle scales $\propto t^2$ in the ballistic phase, $\propto t$ in the diffusive phase, and $\propto$ const. in the localized phase, where $t$ is time \cite{trithep}. We can also characterize these phases according to their inverse participation ratios (IPRs), which we define in real space as $\text{IPR}_{\br} = \sum_{\br} |\psi(\br)|^4$, and in momentum space as  $\text{IPR}_{\bk} = \sum_{\bk} |\psi(\bk)|^4$ where $\psi(\br)$ and $\psi(\bk)$ are energy eigenstates of $H$ in the position and momentum space basis respectively. In the localized phase, $\text{IPR}_{\br} \rightarrow 1$ and $\text{IPR}_{\bk} \rightarrow 0$ in the $L\to\infty$ limit, while in the ballistic phase $\text{IPR}_{\br} \rightarrow 0$ and $\text{IPR}_{\bk} \rightarrow 1$. In the diffusive phase, $\text{IPR}_{\br} \rightarrow 0$ and $\text{IPR}_{\bk} \rightarrow 0$. We plot the IPRs for Eq.~\ref{hamiltonian} as a function of $v$ in Fig.~\ref{fig:IPR}. We estimate the value of $v_{c1}$, the critical point between the ballistic and diffusive phases, to be $\approx 0.5$ for the states near half-filling, as indicated by the dashed vertical line. We also calculate the IPRs for a corresponding 3D system with random disorder. Here, the random disorder potential is realized by making the phases $\phi_i$ fluctuate randomly with position: $\phi_i \rightarrow \phi_i(\br)$, with $\phi_i \in [0, 2\pi]$ chosen randomly on each site. For random disorder, the system is in the diffusive phase for any small $v>0$. 

\section{COOPER LOGARITHM IN QUASIPERIODIC SYSTEMS}
Bardeen-Cooper-Schrieffer (BCS) theory tells us that, in a clean system, the pairing susceptibility in the spin-singlet channel diverges as $\log(T)$ at low temperatures. Specifically, we define the spin-singlet susceptibility in real space as 
\begin{align}
\chi^\text{s}_{\br\br';\br''\br'''} = \int_{0}^\beta d\tau\Big[ &\langle c_{\br\downarrow}(\tau)c^\dagger_{\br''\downarrow}(0)\rangle \langle c_{\br'\uparrow}(\tau)c^\dagger_{\br'''\uparrow}(0)\rangle \nonumber\\
    +~&\langle c_{\br\uparrow}(\tau)c^\dagger_{\br'''\uparrow}(0) \rangle \langle c_{\br'\downarrow}(\tau)c^\dagger_{\br''\downarrow}(0) \rangle\Big],
\end{align} 
where $c_{\br\sigma}(\tau)$ is the electron creation operator at imaginary time $\tau$, and $\beta = 1/T$. If we treat $\chi^\text{s}$ as a $L^6 \times L^6$ matrix, then its largest eigenvalue will scale $\propto \log(T)$ at low temperatures. This is the celebrated Cooper logarithm.
For a system with interactions of the form $H_{\text{int}} = \sum_{\br,\br'} V_{\br\br'} n_{\br} n_{\br'}$, the real-space linearized gap equation (LGE) can be expressed in terms of the pairing susceptibility in the spin-singlet channel as 
\begin{align}
\Delta^\text{s}_{\br\br'} = - \sum_{\br'' , \br'''} V_{\br,\br'} \chi^\text{s}_{\br\br';\br''\br'''} \Delta^\text{s}_{\br''\br'''}.
\end{align}
Here, $\Delta^\text{s}_{\br\br'}$ is the gap function for a pair of particles at $\br$ and $\br'$. The $\log(T)$ divergence of $\chi^{\text{s}}$ therefore guarantees a finite-temperature solution to the LGE for arbitrarily weak attractive interactions. 

In the presence of random non-magnetic impurities, $s$-wave superconductors will continue to exhibit this weak-coupling divergence well into the diffusive phase \cite{Anderson, bennemann2008superconductivity}. We contrast this behavior with that of spin-triplet ($p$-wave) superconductors. The pairing susceptibility for the $S_z=0$ total spin configuration is
\begin{align}
    \chi^\text{p}_{\br\br';\br''\br'''} = \int_{0}^\beta d\tau\Big[ &\langle c_{\br\downarrow}(\tau)c^\dagger_{\br''\downarrow}(0)\rangle \langle c_{\br'\uparrow}(\tau)c^\dagger_{\br'''\uparrow}(0)\rangle \nonumber\\
    -~&\langle c_{\br\uparrow}(\tau)c^\dagger_{\br'''\uparrow}(0) \rangle \langle c_{\br'\downarrow}(\tau)c^\dagger_{\br''\downarrow}(0) \rangle\Big]
\end{align} 
In the case of random disorder, this susceptibility will not diverge for any finite amount of random disorder \cite{sigrist1991phenomenological}.

In this section, we seek to understand how the $s$-wave and $p$-wave susceptibilities behave in the 3D AA model ($p$-wave pairing is chosen because the only well-defined spatial symmetry of the Hamiltonian in Eq. \ref{hamiltonian} is inversion symmetry). To this end, consider the uniform static pairing susceptibility, 
\begin{align}
\chi^{\text{s}/\text{p}, 0}_{\bm{\delta},\bm{\delta'}} = \frac{1}{L^{6}}\sum_{\br,\br'}
\chi^{\text{s}/\text{p}}_{\br,\br+\bm{\delta};\br',\br'+\bm{\delta}'}
\label{eqn:uniform_sus}
\end{align}
Although we do not expect the solution to be spatially uniform for moderate-to-large strengths of the quasiperiodic disorder, $\chi_{\bdelta,\bdelta'}^{\text{s}/\text{p}, 0}$ will nevertheless faithfully capture any divergences present in the actual susceptibility so long as the two quantities possess finite overlap. The spatial average also neutralizes spurious contributions from rare superconducting regions \cite{john}, as the overlap of the uniform solution with any localized solutions goes to zero as $L$ grows large.

\begin{figure}[t]
\centering
\includegraphics[width=0.5\textwidth]{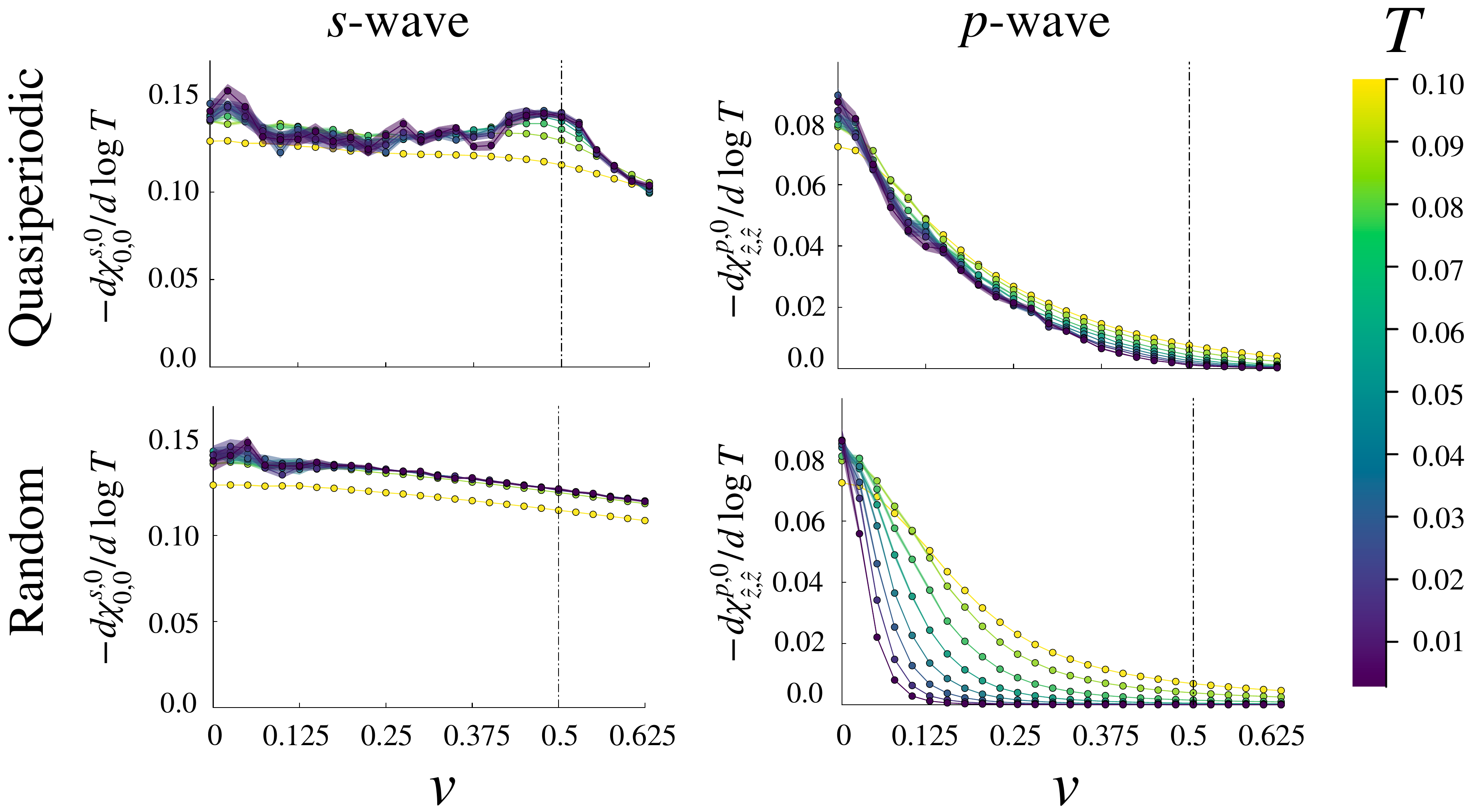}
\caption{The negative derivative of the uniform susceptibilities for quasiperiodic (left) and random (disorder) potentials on a $23\times23\times23$ lattice with $\theta=\pi/7$ and $Q=(\sqrt{2}-1)/2$ along the constant line-cut $0.75=\mu/(1+v)$. The results are averaged over $N\approx 35$ realizations of the potential. The vertical dashed line indicates the ballistic-to-diffusive transition at $v = v_{c1}$. Error ribbons represent the standard error of the mean, which comes from averaging over disorder realizations.}
\label{fig:uniform_sus}
\end{figure}

Assuming pairing of time-reversed states, we calculate the uniform susceptibility \footnote{Taking other components of the susceptibility matrix, such as $\chi^0_{\hat{x},\hat{x}}$ or $\chi^0_{\hat{y},\hat{y}}$, has a negligible quantitative and no qualitative effect on the results for the spin-singlet nearest neighbour and $p$-wave configurations.} for the $s$-wave ($\chi^{\text{s}0}_{0,0}$) and $p_z$-wave  ($\chi^{\text{p}0}_{\hat{z},\hat{z}}$) channels using Eq.~\ref{hamiltonian} with a chemical potential of $\mu = 0.75 (1+v)$. Here we keep $\mu/(1+v)$ constant to account for the enlargement of the bandwidth as $v$ increases. Numerical results for both quasiperiodic and randomly-disordered potentials are shown in Figure \ref{fig:uniform_sus}. For the randomly-disordered potential, we average over different random configurations. For the quasiperiodic potential, we average over different (constant) values of the phases $\{\phi_i \}$. This averaging is essential for capturing the behavior of system sizes larger than what can be simulated. It is worth noting that the averaged system possesses inversion symmetry in both cases. Further details of this calculation are provided in Appendix \ref{appendix:uniform_sus}.

We plot $-d\chi_{\bdelta\bdelta'}^{\text{s}/\text{p}, 0}/d\log T$, the derivative of the uniform susceptibility with respect to the temperature. If this quantity goes to zero, the susceptibility saturates, and superconductivity will therefore exist only in the presence of finite strength interactions. Conversely, a plateau where $-d\chi_{\bdelta\bdelta'}^{\text{s}/\text{p}, 0}/d\log T \neq 0$ is indicative of a $\log(T)$ divergence of the susceptibility at low temperatures---and, by extension, a weak-coupling superconducting instability. We find that the logarithmic divergence of the $s$-wave channel susceptibility is stable to both the quasiperiodic and random potentials. In the $p_z$-wave channel, we find that susceptibility has a logarithmic divergence for $v < v_{c1} \approx .5$. Hence, there is a weak coupling instability in the $p_z$-wave channel throughout the ballistic phase of the model. However, in the diffusive and localized phases ($v > v_{c1}$) this logarithmic divergence is completely suppressed. Nevertheless, we emphasize that the $\log(T)$ divergence of the susceptibility in both conventional and unconventional channels is stable for small but finite quasiperiodicity. This is unlike the case for random disorder, where a finite amount of disorder suppresses the logarithmic divergence in the $p_z$-wave channel. 

\section{TRANSITION TEMPERATURE}\label{section:temp}
Having established the qualitative effects of quasiperiodic disorder on weakly-coupled superconductors, we shift our attention to a concrete interacting model. Here, we consider adding the following interactions to the 3D AA model,
\begin{align}
    H_\text{int} = &V_0\sum_{\br} n_{\br} n_{\br} + \frac{V_1}{2} \sum_{\langle \br,\br'\rangle}  n_{\br} n_{\br'},
    \label{Hint}
\end{align}
where $\langle...\rangle$ indicates nearest neighbours. We consider both $s$-wave and $p$-wave pairing channels. For the $s$-wave channel, we set $V_0 < 0$ and $V_1 =0$, and for $p$-wave, we set $V_0 = 0$ and $V_1 < 0$. We treat the problem at a mean field level by using the following trial Hamiltonians and self-consistency equations:
\begin{align}
    H_\text{MF}^{\text{s}} = \frac{1}{2}&\sum_{\bm{r}} \left[\Delta^{\text{s}}_{\bm{r}\bm{r}} \Big(c^\dagger_{\bm{r}\uparrow} c^\dagger_{\bm{r}\downarrow} -c^\dagger_{\bm{r}\downarrow} c^\dagger_{\bm{r}\uparrow} \Big)+h.c.\right],
\end{align}
\begin{equation}\begin{split}
&\Delta_{\bm{r},\bm{r}}^{\text{s}} = \frac{V_{0}}{2} \Big[ \langle  c_{\bm{r},\downarrow} c_{\bm{r},\uparrow} \rangle - \langle c_{\bm{r},\uparrow} c_{\bm{r},\downarrow} \rangle\Big],
\label{eq:selfCS}\end{split}\end{equation}
and 
\begin{align}
    H_\text{MF}^\text{p} = \frac{1}{2}\sum_{\langle \bm{r},\bm{r'}\rangle} \left[\Delta^\text{p}_{\bm{r}\bm{r'}} \Big(c^\dagger_{\bm{r}\uparrow} c^\dagger_{\bm{r'}\downarrow} +c^\dagger_{\bm{r}\downarrow} c^\dagger_{\bm{r'}\uparrow} \Big)+h.c.\right],
\end{align}
\begin{equation}
\Delta_{\bm{r},\bm{r}'}^{\text{p}} = \frac{V_{1}}{2} \Big[ \langle  c_{\bm{r'},\downarrow} c_{\bm{r},\uparrow} \rangle + \langle c_{\bm{r'},\uparrow} c_{\bm{r},\downarrow} \rangle\Big],
\label{eq:selfCP}\end{equation}
for $s$- and $p$-wave respectively. 

For both pairing symmetries, the mean-field transition temperature can be identified by studying solutions of the linearized gap equation. Given an interaction matrix $V_{\br\br'}$ defined as in Eq. \ref{Hint}), we estimate $T_c^\text{LGE}$ by diagonalising
\begin{equation}
    M^{\text{s}/\text{p}}_{\br\br';\br''\br'''}\equiv - \sum_{\br'' , \br'''} V_{\br\br'} \chi^{\text{s}/\text{p}}_{\br\br';\br''\br'''}
\end{equation}
and determining the temperature at which the largest eigenvalue of $M^{\text{s}/\text{p}}$ is equal to $1$ \footnote{Here $\chi^{\text{s}/\text{p}}$ is the susceptibility calculated in the normal state ($\Delta^{\text{s}/\text{p}} = 0$)}. 
\begin{figure}[t]
\centering
\includegraphics[width=0.5\textwidth]{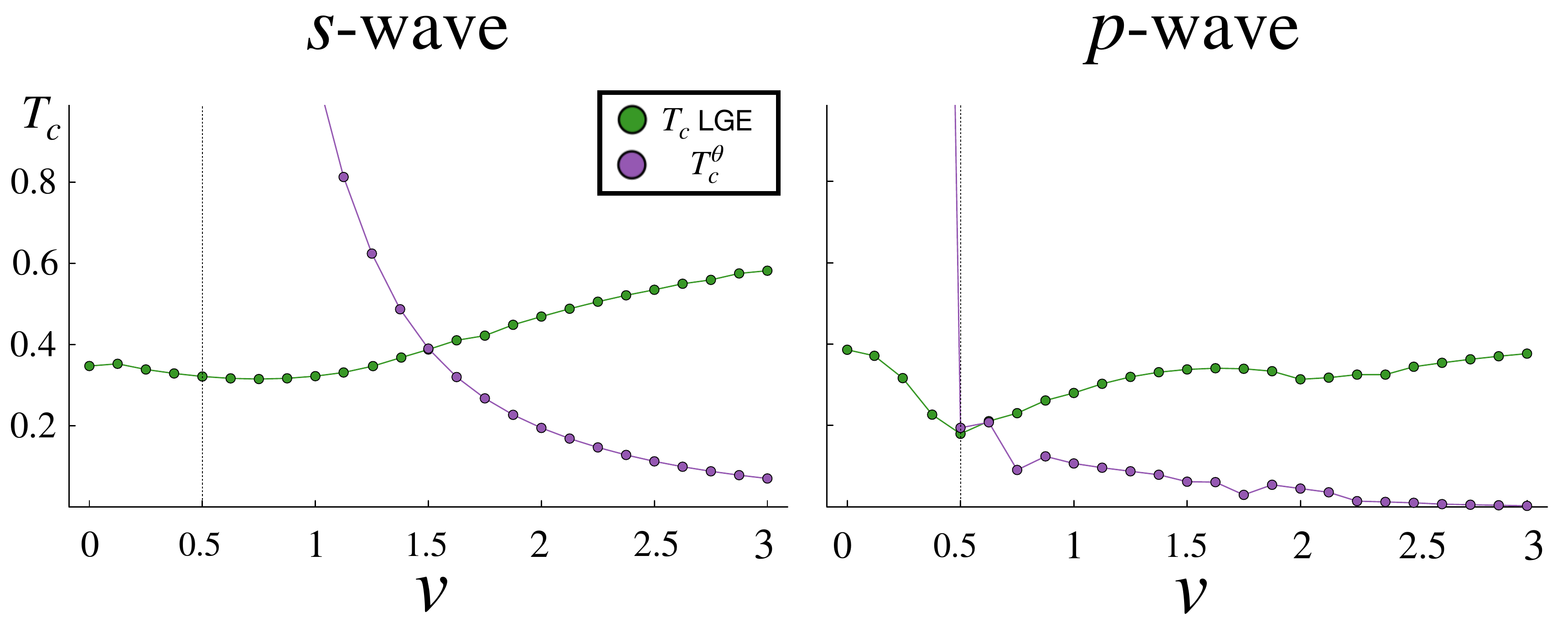}
\caption{The transition temperatures calculated for $s$-wave ($V_0=-3,V_1=0$) and $p$-wave ($V_0=0, V_1=-3)$ superconductivity for a $11\times 11\times 11$ lattice. The green markers correspond to $T_c^\text{LGE}$, the transition temperature calculated from the linearized gap equation. The purple markers correspond to $T^\theta_c$, the transition temperature as estimated from the superfluid stiffness.}
\label{fig:Tc}
\end{figure}

The LGE solution is plotted using green markers in Figure \ref{fig:Tc}, for $\mu = 0.75(1+v)$. For both $s-$ and $p-$wave systems, we observe an initial decrease followed by an increase in $T_c^\text{LGE}$ as we increase $v$. This latter increase in $T_c^\text{LGE}$ is unphysical, in the sense that it does not correspond to the transition into a state with zero resistance \cite{john}. To understand why the LGE solution behaves in this way, it is important to keep in mind that $T_c^\text{LGE}$ estimates the temperature at which a superconducting condensate first forms. However, this condensate may be localized to multiple small, non-overlapping regions. In this setting, one should really think of the system as being composed of superconducting islands coupled to each other via weak Josephson-type coupling \cite{stiffness}. The resistance of such a state will only drop to zero when the different islands become phase coherent. If the coupling between different islands is weak, the phases might not cohere until the temperature is lowered \textit{below} $T_c^\text{LGE}$.

To estimate the temperature at which phase coherence sets in, we assume that the only relevant degrees of freedom are the superconducting phases, and model the superconductor as an XY-model. The temperature where phase coherence sets in is estimated as  $T^\theta_c = \frac{An_s(0)}{4m^*}  a_\text{SC} $ \cite{stiffness}, where $n_s(0)$ is the geometric mean of the super fluid density at zero temperature \cite{superfluid_density}\footnote{Owing to the anisotropy of the model, the superfluid density need not be the same along every dimension. In practice, however, we find that the densities along each spatial dimension are roughly equal, and therefore appropriately estimated by the geometric mean}, $m^*$ is the effective mass of the electrons, $a_\text{SC}=\min\{\sqrt{\pi}\xi_\text{SC},a_\text{L}\}$; $a_\text{L}$ is the lattice constant, here taken to be $1$ for convenience; $\xi$ is the superconducting coherence length, $\xi = v_F/(\pi \Delta)$\footnote{Because $\Delta$ varies with position, we take the maximum value of $\Delta_{\br,\br'}$.}; and $v_F$ is the fermi velocity calculated in the clean system. $A$ is a dimensionless number of order unity that depends on the details of the short-distance physics, which we take here to be $2.2$ after Ref.~\cite{stiffness}. The superfluid stiffness $n_s$ and coherence length $\xi$ are calculated by solving the BdG equations (Eq.~\ref{eq:selfCS} and \ref{eq:selfCP}); see Appendices \ref{appendix:BdG} and \ref{appendix:stiffness} for more details.

Numerical results for the transition temperatures calculated from the LGE solution and the superfluid stiffness are shown in Fig~\ref{fig:Tc}. The true $T_c$ of the system is $T_c = \min(T_c^\text{LGE}, T_c^\theta )$. For $s$-wave $T_c$ remains finite even into the localized phase $v > v_{c2}$. Conversely, $T_c$ falls off sharply close to $v_{c1}$ for the $p$-wave case. At large $v$, $T^{\text{LGE}}_c$ continues to artificially increase for both the $s$- and $p$-wave pairing symmetries on account of contributions from rare superconducting regions. However, $T^\theta_c$ falls off quickly at large $v$, leading to an expected decrease of the superconducting transition temperature when the quasiperiodic potential dominates. These results agree with our previous analysis of the susceptibilities. Namely, we find that $p$-wave superconductivity is robust in the ballistic phase but is suppressed in the diffusive phase, while $s$-wave remains robust well into the localized phase. 

\section{CONCLUSIONS}
\label{sec:conclusion}
Having described the effects of quasiperiodicity on $s$- and $p$-wave superconductivity, one might naturally wonder whether these results generalize to other pairing symmetries as well. Owing to a lack of rotation symmetry in the model, there is no sharp distinction between, for example, $s$- and $d$-wave pairing symmetries. Nevertheless, in Appendix \ref{appendix:dwave}, we present results on spin-singlet nearest neighbour pairing, which reduces to the $d_{x^2-y^2}$ pairing term in the $v\to 0$ limit. We find good qualitative agreement between the $p$-wave and nearest-neighbour spin-singlet data, suggesting that the essential element resulting in the suppression of the log divergence is not the pairing symmetry, but rather whether the gap integrates to zero over the Brillouin zone.  

In summary, we have shown that quasiperiodicity behaves quite unlike random disorder in the context of unconventional superconductivity. We find that $p$-wave superconductivity is robust up to a finite critical value of the potential strength $v_{c1}$. This value corresponds to the ballistic-to-diffusive transition of the non-interacting model. $s$-wave superconductivity, conversely, survives in all three phases---ballistic, diffusive, and localized. 
Our results do not rule out the possibility of unconventional superconductivity in quasiperiodic twisted trilayer graphene \cite{aviram}; however, further theoretical studies in 2D and including microscopic details are necessary to fully explore this possibility. 

\begin{acknowledgments}
We thank Steve Kivelson, Srinivas Raghu, John Dodaro, Gil Refael, and Rafael Fernandes for insightful discussions. This work was supported in part by a startup fund at Stanford University. EB and JX were supported by the Israel-USA Binational Science Foundation (BSF) and the European Research Council (ERC) under grant HQMAT (Grant Agreement No. 817799).
\end{acknowledgments}

\bibliography{quasiperiodic}

\appendix
\beginsupplement
\pagebreak
\section{Model characterization}\label{appendix:IPR}
\begin{figure}
\centering
\includegraphics[width=0.75\textwidth]{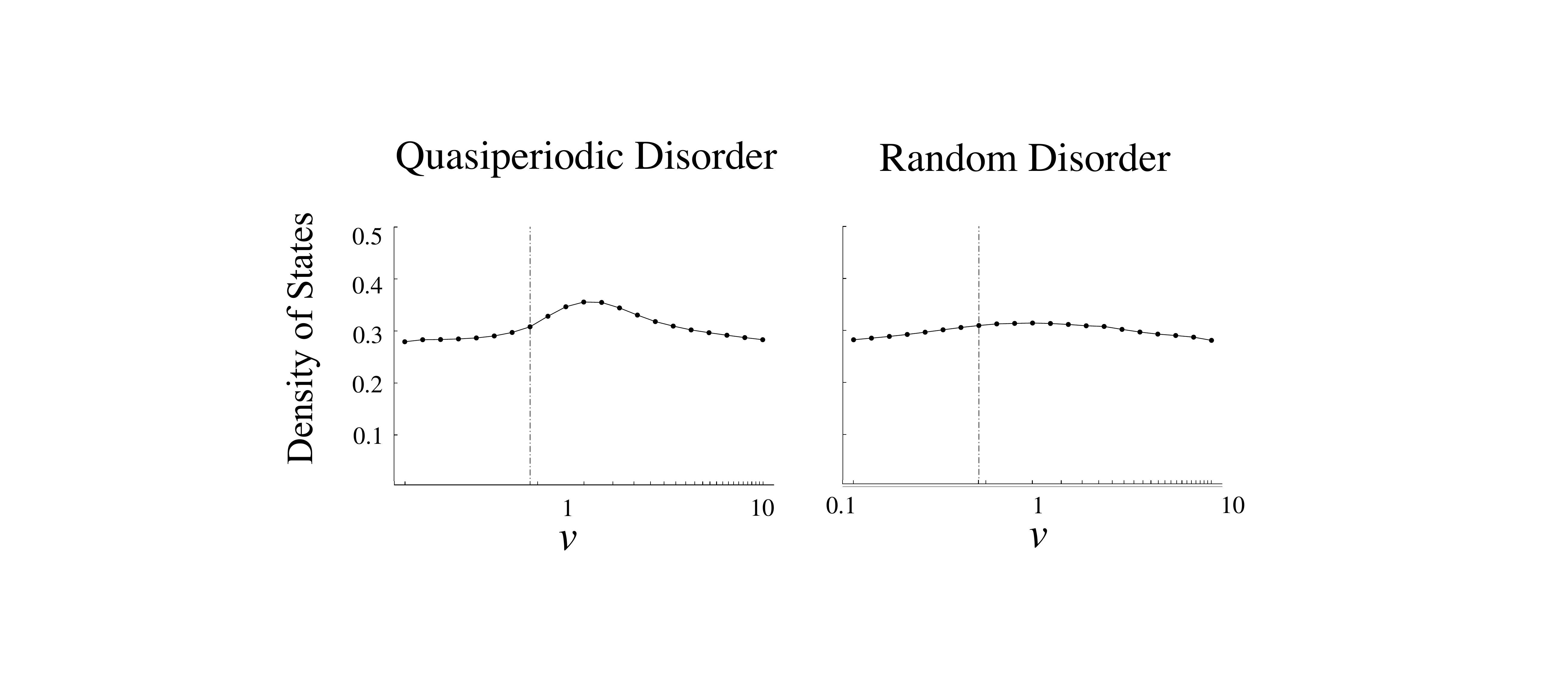}
\caption{Density of states as a function of disorder strength $v$ along the $0.75=\mu/(1+v)$ line cut. The location of the ballistic-to-diffusive phase transition is marked by a vertical dashed line. Results are calculated for a $23\times 23\times 23$ lattice with $\theta=\pi/7$ and $Q=(\sqrt{2}-1)/2$.}
\label{fig:IPR_appendix}
\end{figure}

In Figure \ref{fig:IPR_appendix}, we show the density of states along the $0.75=\mu/(1+v)$ line cut, the chemical potential at which all other data were collected for this paper. We estimate the location of $v_{c1}$, the ballistic-to-diffusive phase transition, as the point where the momentum-space IPR drops to near-zero (see main text). This point is indicated by a vertical dashed line. Data are shown for both quasiperiodic and random disorder.
\section{Nearest-neighbour pairing}\label{appendix:dwave}
\begin{figure}
\centering
\includegraphics[width=0.95\textwidth]{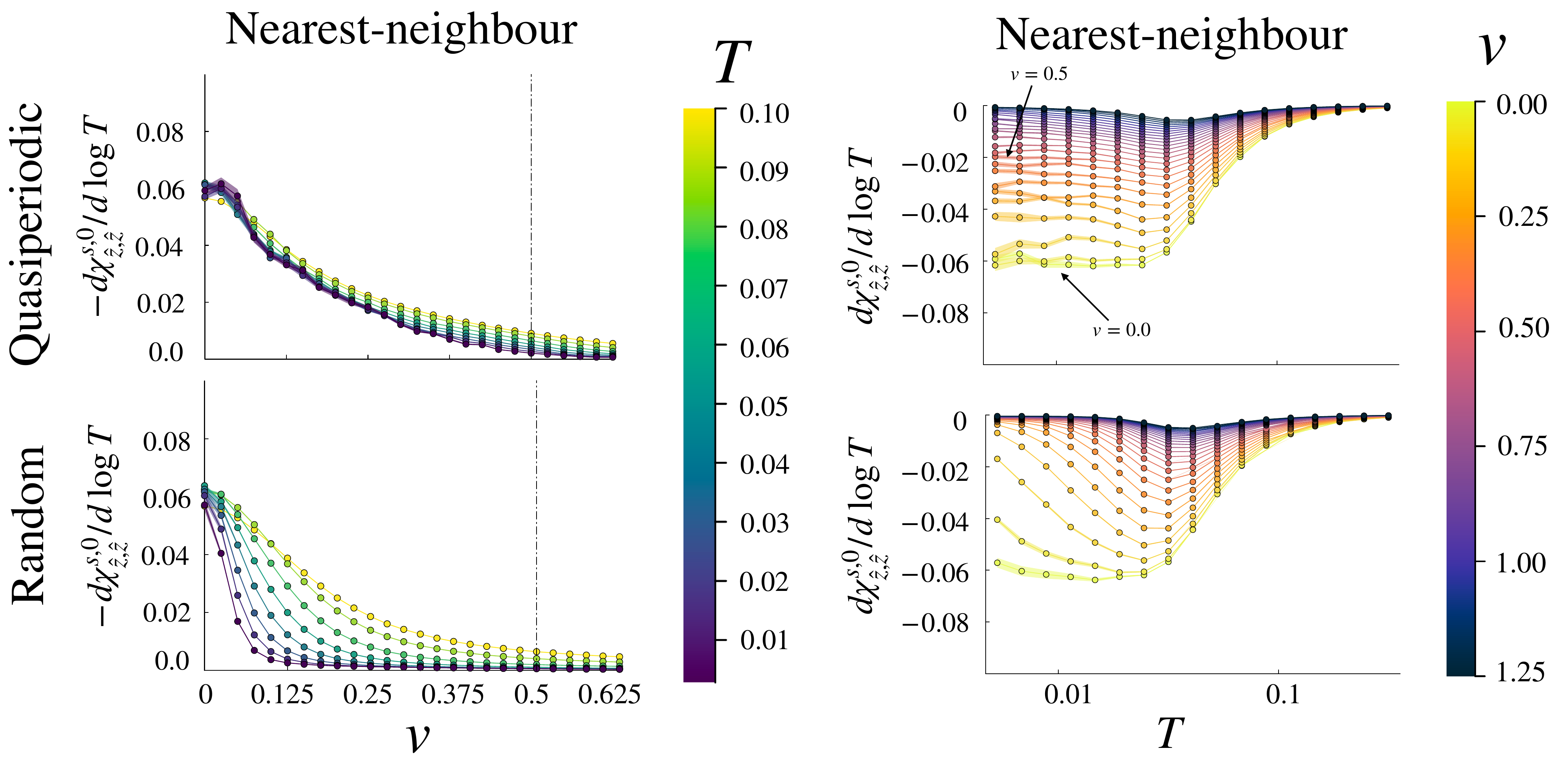}
\caption{We show $-d\chi^0_{\bdelta,\bdelta'}/d\log T$ as a function of disorder strength (left column) and $d\chi^0_{\bdelta,\bdelta'}/d\log T$ as a function of $T$ (right column) in the case of spin-singlet nearest-neighbour pairing. These results are calculated for both quasiperiodic (top row) and random (bottom row) disorder potentials on a $23\times 23\times 23$ lattice with $\theta=\pi/7$, $Q=(\sqrt{2}-1)/2$, and $0.75=\mu/(1+v)$. Results are averaged over $N\approx 35$ disorder realizations with error ribbons showing the standard error of the mean.}
\label{fig:dwave_appendix}
\end{figure}

In Figure \ref{fig:dwave_appendix}, we repeat the analyses from the main text in the case of nearest-neighbour spin-singlet pairing. We treat the problem at a mean field level by using the following trial Hamiltonians and self-consistency equations:
\begin{align}
    H_\text{MF}^{\text{s,nn}} = \frac{1}{2}&\sum_{\langle \bm{r},\bm{r'}\rangle} \Delta^{\text{s,nn}}_{\bm{r}\bm{r'}} \Big(c^\dagger_{\bm{r}\uparrow} c^\dagger_{\bm{r}'\downarrow} -c^\dagger_{\bm{r}\downarrow} c^\dagger_{\bm{r}'\uparrow} \Big),
\end{align}
\begin{equation}\begin{split}
&\Delta_{\bm{r},\bm{r}'}^{\text{s,nn}} = \frac{V_{1}}{2} \Big[ \langle  c_{\bm{r}',\downarrow} c_{\bm{r},\uparrow} \rangle - \langle c_{\bm{r}',\uparrow} c_{\bm{r},\downarrow} \rangle\Big],
\end{split}\end{equation}
\begin{figure}
\centering
\includegraphics[width=0.4\textwidth]{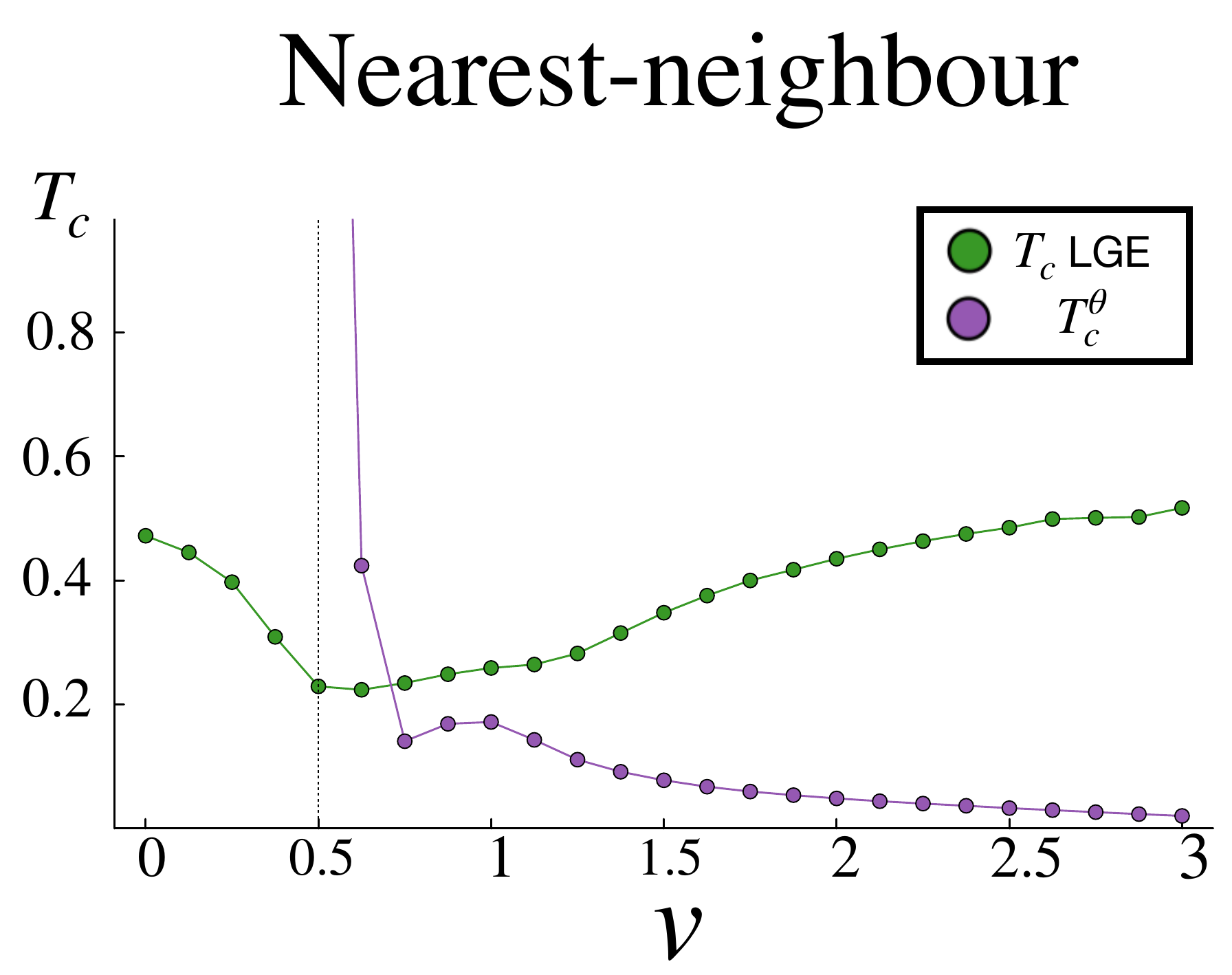}
\caption{Transition temperature estimates for the case of nearest-neighbour spin-singlet pairing. Green line represents the LGE estimate, purple corresponds to superfluid stiffness estimate, and the black dashed line represents the minimum of the two. These data are calculated on an $11\times 11\times 11$ lattice with $V_0=1,V_1=-3$ interaction strength at $\theta=\pi/7$, $Q=(\sqrt{2}-1)/2$, and $0.75=\mu/(1+v)$.}
\label{fig:dwave_appendix_Tc}
\end{figure}
\section{BdG Hamiltonian and Self-Consistency Equations} \label{appendix:BdG}
We can write the lattice Hamiltonian of a general superconducting system as
\begin{equation}
    H = -\sum_{\br\br';\alpha}h^{\alpha}_{\br\br'}c^\dagger_{\br\alpha}c_{\br'\alpha}+V_\text{int},
\end{equation}
(We now use $\alpha$ to denote spin, to eliminate confusion in subsequent parts of this derivation). The interaction term can be decomposed into singlet and triplet parts:
\begin{equation}
    V_\text{int} = \frac{1}{2}\sum_{\br\br'}\sum_{\alpha_1...\alpha_4}\sum_{m=0}^3 g^m_{\br'\br}(\sigma_m i \sigma_2)^\dagger_{\alpha_3\alpha_1}(\sigma_m i \sigma_2)_{\alpha_2\alpha_4}(c_{\br\alpha_1}c_{\br'\alpha_3})^\dagger c_{\br\alpha_2}c_{\br'\alpha_4}\label{Vpair}
\end{equation}
The tensor $g^m_{\br'\br}$ contains information about the pairing symmetry; $g_{0\br'\br}$ is the singlet component, while $g_{1\br'\br},g_{2\br'\br},g_{3\br'\br}$ are the triplet parts. For example, the following choices will give us the pairing symmetries considered in this paper:
\begin{itemize}
    \item $g_{\br\br}^0=V_0$ and  $g_{\br\neq\br'}^0=g_{\br\br'}^{m\neq 0}=0$ gives $s$-wave pairing
    \item $g_{\br\br}^0=V_0$, $g_{\langle\br\br'\rangle}^0=V_1$, and $g_{\br\br'}^{m\neq 0}=0$ gives spin-singlet nearest-neighbour pairing
    \item $g_{\br\br'}^{0}=g_{\br\br'}^{1}=g_{\br\br'}^{2}=0$ and $g_{\br\br'}^{3}=V_1$ yields spin-triplet pairing of type $\left(|\up\down\rangle+|\down\up\rangle\right)/\sqrt{2}$
\end{itemize}
We now perform a mean field decomposition on Eq.~\ref{Vpair}:
\begin{equation}
    V_\text{int} = \frac{1}{2}\sum_{\br\br'}\sum_{\alpha_1...\alpha_4}\sum_{m=0}^3 g^m_{\br'\br}(\sigma_m i \sigma_2)^\dagger_{\alpha_3\alpha_1}(\sigma_m i \sigma_2)_{\alpha_2\alpha_4}\left[\langle c^\dagger_{\br\alpha_1}c^\dagger_{\br'\alpha_3}\rangle c_{\br\alpha_2}c_{\br'\alpha_4} +  c^\dagger_{\br\alpha_1}c^\dagger_{\br'\alpha_3} \langle c_{\br\alpha_2}c_{\br'\alpha_4}\rangle\right] 
\end{equation}
We can define the order parameter as follows:
\begin{equation}
    \begin{aligned}
    \Delta_{\br'\br,\alpha_2\alpha_4} &= \sum_{m=0}^3 d^m_{\br'\br}(\sigma_m i \sigma_2)_{\alpha_2\alpha_4}\\
    &= \begin{pmatrix}
-d_x(\bk)+i d_y(\bk) & d_z(\bk) + \Delta_s(\bk) \\
d_z(\bk)-\Delta_s(\bk) & d_x(\bk)+id_y(\bk) 
\end{pmatrix}
\end{aligned}
\end{equation}
where 
\begin{equation}
d^m_{\br'\br}=g^m_{\br'\br}\sum_{\alpha_1\alpha_3}(\sigma_m i \sigma_2)_{\alpha_3\alpha_1}\langle c_{\br\alpha_1}c_{\br'\alpha_3}\rangle
\end{equation}
In more transparent notation,
\begin{align}
    d^0 \equiv \Delta_s &= \frac{1}{2}\left(\Delta^{+-}-\Delta^{-+}\right)
\end{align}
This is the spin-singlet order parameter, where $+(-)$ is spin up (down). The spin-triplet order parameter, conversely, is a vector, and it is given by
\begin{align}
    d^{m>0} \equiv \mathbf{d} &= \frac{1}{2}\left(\Delta^{--}-\Delta^{++},-i(\Delta^{--}+\Delta^{++}),\Delta^{+-}+\Delta^{-+} \right)
\end{align}
The self-consistency equation for mean-field spin-singlet $s$-wave is therefore:
\begin{equation}
    \Delta_{\br\br}^{\text{s},s} = \frac{V_0}{2}\Big(\langle c_{\br\uparrow}c_{\br\downarrow}\rangle - \langle c_{\br\downarrow}c_{\br\uparrow} \rangle\Big) \label{SC_s}
\end{equation}
For spin-singlet nearest-neighbour pairing, we have:
\begin{equation}
\Delta_{\br\br'}^{\text{s},\text{nn}} = \frac{V_1}{2}\Big(\langle c_{\br'\uparrow}c_{\br\downarrow} \rangle- \langle c_{\br'\downarrow}c_{\br\uparrow} \rangle\Big) \label{SC_nn}
\end{equation}
And, finally, for $p$-wave spin-triplet pairing, the order parameter is given by 
\begin{equation}
\Delta_{\br\br'}^{\text{p},p} = \frac{V_1}{2}\Big(\langle c_{\br'\uparrow}c_{\br\downarrow} \rangle+ \langle c_{\br'\downarrow}c_{\br\uparrow} \rangle\Big) \label{SC_p}
\end{equation}
We diagonalize equation \ref{Vpair} by introducing the Bogoliubov quasiparticle operators $(\gamma_n,\gamma_n^\dagger)$: 
\begin{align}
    \begin{pmatrix}
c_{\br\uparrow} \\
c^\dagger_{\br\downarrow}\end{pmatrix} &= \sum_n \begin{pmatrix}
u_{\br n\up} & v^*_{\br n\up} \\
v_{\br n\down} & u^*_{\br n\down}\end{pmatrix} \begin{pmatrix}
\gamma_{\br\uparrow} \\
\gamma^\dagger_{\br\downarrow}\end{pmatrix} \label{transform}
\end{align}
In this basis, \ref{Vpair} is diagonalized, with eigenequations given by
\begin{align}
    \sum_{j}\begin{pmatrix}
\mathcal{H}_{\br\br'} & \Delta_{\br\br'} \\
\Delta^\dagger_{\br'\br} & -\mathcal{H}^*_{\br\br'} 
\end{pmatrix}\begin{pmatrix}
u_{\br'\uparrow} \\
v_{\br'\downarrow}
\end{pmatrix} = E_n \begin{pmatrix}
u_{\br\uparrow} \\
v_{\br\downarrow}
\end{pmatrix}
\end{align}
We now have three sets of self-consisten equations, one set for each of the pairing symmetries we are interested in studying here. By substituting Eq.~\ref{transform} into \ref{SC_s}, \ref{SC_nn}, and \ref{SC_p}, we can derive expressions for the order parameters with respect to the Bogoliubov coefficients:
\begin{equation}
    \Delta_{\br\br'} = \frac{V_{\br\br'}}{4}\sum_n (u_{\br\uparrow}^n v_{\br'\downarrow}^{n*} + h.c.)\tanh{\left(\frac{E_n}{k_BT}\right)}
\end{equation}
for spin-singlet pairing, and 
\begin{equation}
    \Delta_{\br\br'} = \frac{V_{\br\br'}}{2}\sum_n \Big[u_{\br\uparrow}^n v_{\br'\downarrow}^{n*}f(-E_n) + u_{\br'\uparrow}^n v_{\br\downarrow}^{n*}f(E_n)\Big]
\end{equation}
for spin-triplet ($p$-wave) superconductivity. In the above expression, $f(E_n)$ is the Fermi-Dirac distribution evaluated at energy $E_n$.
\section{Linearized Gap Equation}
We are interested in calculating the pair susceptibilities,
\begin{align}
    \chi_{ab;cd} &= \int_{0}^\beta d \tau\langle \psi_{ab}(\tau)\psi^\dagger_{cd}(0)\rangle\label{susc}
\end{align}
for the spin-singlet and triplet configurations, with the singlet operator defined as
\begin{equation}
    \psi^\text{s}_{ab} = \frac{1}{2}\Big(\langle c_{a\uparrow}c_{b\downarrow}\rangle - \langle c_{a\downarrow}c_{b\uparrow} \rangle\Big)
\end{equation}
and triplet operator
\begin{equation}
    \psi^\text{p}_{ab} = \frac{1}{2}\Big(\langle c_{a\uparrow}c_{b\downarrow}\rangle + \langle c_{a\downarrow}c_{b\uparrow} \rangle\Big)
\end{equation}
Substituting these equations for the operators into Eq.~\ref{susc}, we have
\begin{equation}
    \chi_{ab;cd} = \frac{1}{2}\int_0^\tau d\tau\Big[\langle c_{a\down}(\tau)c_{c\down}^\dagger(0) \rangle \langle c_{b\up}(\tau)c_{d\up}^\dagger(0) \rangle \pm \langle c_{a\down}(\tau)c_{d\down}^\dagger(0) \rangle \langle c_{b\up}(\tau)c_{c\up}^\dagger(0) \rangle\Big] \label{susc2}
\end{equation}
The $+$ sign is for spin-singlet pairing, while the $-$ is for spin-triplet. We solve for the pair susceptibility by diagonalizing the non-interacting Hamiltonian with quasiperiodic disorder: $H_0 = \sum_n E_n a^\dagger_n a_n$. Making the substitution $c_{\br\sigma} = \sum_n U^*_{\br n \sigma} a_{n}$, we can calculate the Green's function 
\begin{equation}
    \langle c_{a\sigma}(t) c_{b\lambda}^\dagger(0)\rangle = \sum_n U^*_{a n\sigma} U_{b n\lambda} e^{-\tau E_n}\times [1-f(E_n)]
\end{equation}
Substituting this expression in \ref{susc2}, we can derive an expression for the susceptibility in terms of the eigenstates of the single-particle Hamiltonian:
\begin{equation}
    \chi_{ab;cd}=\frac{1}{2}\sum_{n,m}\left(\frac{1-f(E_n)-f(E_m)}{E_n+E_m}\right)\times\Big(U^*_{a n \down}U_{c n\down}U^*_{b m \up} U_{d m \up} \pm U^*_{a n\up} U_{d n\up} U^*_{b m \down} U_{c m \down}\Big)
\end{equation}
Once again, we remind the reader that the $+$ sign is for spin-singlet superconductivity, while the $-$ sign is reserved for spin-triplet. 

\section{Superfluid stiffness} \label{appendix:stiffness}
The signature of the onset of the superconducting phase is the Meissner effect, in which a sufficiently weak magnetic field is expelled from the bulk of the superconductor up to a finite penetration depth, $\lambda$. The penetration depth is inversely proportional to the superfluid stiffness. Therefore, the superfluid stiffness and its temperature dependence set a criterion for superconductivity. In this section, we will derive a formula for the superfluid stiffness in the basis of the BdG coefficients following the derivation in \cite{BdG_book}. In the presence of an electric field applied in the $\hat{x}$ direction, the system will respond with a current,
\begin{equation}
    J_{x}(\br_i,t)=\sigma_{xx}(\br_i,\omega)E_x(\br_i,t),
\end{equation}
where $\sigma_{xx}(\mathbf{r}_i,\omega)$ is the local conductivity at the $i$th site. We will work in the Coulomb gauge, where the potential can be expressed as
\begin{equation}
    A_x(\br,t) = -\frac{i}{\omega}E_x(\br,t) = -\frac{i}{\omega} \Xi_x e^{i(\bq\cdot\br_i-\omega t)}
\end{equation}
In the presence of this gauge field, the hopping elements pick up a complex exponential factor:
\begin{equation}
    -\sum_{ ij,\sigma}t_{ij}c^\dagger_{i,\sigma}c_{j,\sigma}  \to -\sum_{ ij,\sigma}\Tilde{t}_{ij}c^\dagger_{i,\sigma}c_{j,\sigma}  
\end{equation}
where $\Tilde{t} = t_{ij}e^{i\phi_{ij}(t)}$, for gauge phase $\phi_{ij}(t) \equiv eA_{ij} = eA_x(\br_i,t)(x_i-x_j)$. Expanding $\Tilde{t}$ up to second order in $A$, we have
\begin{equation}
    -\sum_{ ij,\sigma}\Tilde{t}_{ij}c^\dagger_{i,\sigma}c_{j,\sigma}  \approx -\sum_{ij,\sigma}t\left[1+i\phi_{ij}-\frac{1}{2}\phi_{ij}^2 \right]c^\dagger_{i,\sigma}c_{j,\sigma} 
\end{equation}
Moving forward, we restrict the sum to next nearest neighbours on a lattice with spacing $a$. The new Hamiltonian, in the presence of the gauge field $A$, can be written in terms as
\begin{equation}
    \begin{aligned}
    H &= H_0 + H_\text{int} -it \sum_{\langle i,j\rangle, \sigma}\left[\phi_{ij}(t) c^\dagger_{i,\sigma}c_{j,\sigma} + h.c.\right] + \frac{1}{2}t\sum_{\langle i,j\rangle, \sigma}\left[\phi^2_{ij}(t)c^\dagger_{i,\sigma}c_{j,\sigma} + h.c.\right]\\
    &= H_0 + H_\text{int} - iea \sum_{i,\sigma}\left[A_x(\br_{i+\hat{x}},t)c^\dagger_{i+\hat{x},\sigma}c_{i,\sigma}-A_x(\br_i,t)c^\dagger_i c_{i+\hat{x}}\right]\\
    &\quad\quad+\frac{e^2a^2}{2}\sum_{i,\sigma}\left[A^2_x(\br_{i+\hat{x}},t)c^\dagger_{i+\hat{x},\sigma}c_{i,\sigma}-A^2_x(\br_i,t)c^\dagger_i c_{i+\hat{x}}\right] \nonumber
\end{aligned}
\end{equation}
To simplify this expression, we introduce the paramagnetic current $J_x^P(\br_i,t)$ and the local kinetic energy density $K_x(\br_i,t)$, which are defined as follows:
\begin{align}
    J_x^P(\br_i)&= it\sum_{i,\sigma}\left[c^\dagger_{i+\hat{x},\sigma}c_{i,\sigma}-c^\dagger_{i,\sigma}c_{i+\hat{x},\sigma}\right]\\
    K_x(\br_i)&= -t\sum_{i,\sigma}\left[c^\dagger_{i+\hat{x},\sigma}c_{i,\sigma}+c^\dagger_{i,\sigma}c_{i+\hat{x},\sigma}\right]
\end{align}
Therefore, the Hamiltonian in the presence of the gauge potential can be expressed as 
\begin{equation}
    \begin{aligned}
    H &= H_0 + H_\text{int} -ea \sum_i A_x(\br_i,t)J_x^P(\br_i)-\frac{e^2a^2}{2}\sum_i A_x^2(\br_i,t)K_x(\br_i)\\
    &\equiv H_0 + H_\text{int} + H'(t)
\end{aligned}
\end{equation}
The total $x$-oriented current is given by the sum of the paramagnetic and diamagnetic pieces:
\begin{equation}
    J_x(\br_i,t) = -\frac{\delta H'(t)}{\delta A-x(\br_i,t)}=eaJ_x^P(\br_i)+e^2a^2K_x(\br_i)A_x(\br_i,t)
\end{equation}
Moving forward, we set the lattice constant $a=1$. To determine the superfluid stiffness, we must take the expectation value of the total current:
\begin{equation}
    \begin{aligned}
    \langle J_x(\br_i,t)\rangle &= e\langle J_x^P(\br_i)\rangle - e^2\langle K_x(\br_i)\rangle A_x(\br_i,t) \\
    &\approx ie\int_{-\infty}^t \dd t'\langle[J_x^P(\br_i,t),H^{'P}_I(t')] \rangle - e^2\langle K_x(\br_i)\rangle A_x(\br_i,t)
\end{aligned}
\end{equation}
In going from the first to the second line above, we applied the Kubo formula to find $\langle J_x^P(\br_i,t)\rangle$. Here, 
\begin{align}
    H^{'P}_I(t') = -e\sum_{i,\sigma}A_x(\br_i,t')J_x^P(\br_i,t')
\end{align}
This is the perturbative piece of the Hamiltonian (due just to the paramagnetic term) in the interaction picture. After expressing the current operator in Fourier space,
\begin{equation}
    J_x^P(\bq,t)=\sum_i e^{-i\bq\cdot \br_i}J_x^P(\br,t),
\end{equation}
we can derive
\begin{equation}
    \begin{aligned}
    \langle J_x^P(\br_i,t)\rangle &=\frac{e}{\omega}\Xi_x \int_{-\infty}^t \dd t' \langle [J_x^P(\br_i,t),J_x^P(-\bq,t')]\rangle e^{-i\omega t'}\\
    &= \frac{e}{\omega}\Xi_x e^{-i\omega t} \int_{0}^\infty \dd t'' \langle [J_x^P(\br_i,t''),J_x^P(-\bq,0)]\rangle e^{i\omega t''}\\
    &= \frac{e}{\omega}E_x(\br_i,t)e^{-i\bq\cdot\br_i}  \int_{0}^\infty \dd t'' \langle [J_x^P(\br_i,t''),J_x^P(-\bq,0)]\rangle e^{i\omega t''}
\end{aligned}
\end{equation}
In the second line we made the substitution $t''=t-t'$. The total current density is therefore
\begin{align}
    \langle J_x(\br_i,t)\rangle = \frac{e^2}{\omega}E_x(\br_i,t)e^{-i\bq\cdot\br_i}  \int_{0}^\infty \dd t'' \langle [J_x^P(\br_i,t''),J_x^P(-\bq,0)]\rangle e^{i\omega t''} - \frac{ie^2}{\omega}\langle K_x(\br_i)\rangle E_x(\br_i,t)
\end{align}
To find the local conductivity, we divide by the local electric field:
\begin{equation}
    \begin{aligned}
    \sigma_{xx}(\br_i,\omega) &= \frac{\langle J_x(\br_i,t)\rangle}{E_x(\br_i,t)} \\
    &=\frac{e^2}{\omega}e^{-i\bq\cdot\br_i}  \int_{0}^\infty \dd t'' \langle [J_x^P(\br_i,t''),J_x^P(-\bq,0)]\rangle e^{i\omega t''} - \frac{ie^2}{\omega}\langle K_x(\br_i)\rangle 
\end{aligned}
\end{equation}
To eliminate atomic fluctuations, we average over the spatial variable $\br_i$, yielding
\begin{align}
    \sigma_{xx}(\br_i,\omega)&=\frac{e^2}{N\omega}e^{-i\bq\cdot\br_i}  \int_{0}^\infty \dd t'' \langle [J_x^P(\bq,t''),J_x^P(-\bq,0)]\rangle e^{i\omega t''} - \frac{ie^2}{N\omega}\langle K_x\rangle 
\end{align}
Here, $\langle K_x\rangle = \frac{1}{N}\sum_i \langle K_x(\br_i)\rangle$. We can extend the limits to $\pm \infty$ by introducing a Heaviside multiplier, $\Theta(t'')$:
\begin{equation}
    \begin{aligned}
    \sigma_{xx}(\br_i,\omega)&=\frac{e^2}{N\omega}e^{-i\bq\cdot\br_i}  \int_{-\infty}^\infty \dd t'' \Theta(t'')\langle [J_x^P(\bq,t''),J_x^P(-\bq,0)]\rangle e^{i\omega t''} - \frac{ie^2}{N\omega}\langle K_x\rangle \\
    &= \frac{e^2}{i\omega}\left[-\Pi_{xx}(\bq,\omega) + \langle K_x\rangle \right]
\end{aligned}
\end{equation}
Where we've defined the retarded correlation function of the particle current operator as 
\begin{equation}
    \Pi_{xx}(\bq,t) = -\frac{i}{N}\Theta(t)\langle[J_x^P(\bq,t),J_x^P(-\bq,0)]\rangle, 
\end{equation}
and its corresponding Fourier transform,
\begin{equation}
    \Pi_{xx}(\bq,\omega) = \int_{-\infty}^\infty d t e^{i\omega t}\Pi_{xx}(\bq,t)
\end{equation}
Technically, we calculate the current-current correlator in the Matsubara formalism and then perform the analytic continuation $i\omega_n \to \omega + i\delta$. This means that the conductivity formula is actually given by
\begin{equation}
    \sigma_{xx}(\bq,\omega) = \frac{e^2}{i(\omega+i\delta)}\left[-\Pi_{xx}(\bq,\omega) + \langle K_x\rangle \right]
\end{equation}
The onset of superconductivity is characterized by the Meissner effect, which involves the current response to a static ($\omega =0$) and transverse gauge potential ($\bq\cdot\mathbf{A}=0$). For the Coulomb gauge, $\mathbf{A}$ is entirely along the $\hat{x}$ direction, so we must only satisfy $\bq\cdot\mathbf{A}=q_x A_x=0$. Allowing for arbitrary $\mathbf{A}$, this means that $q_x=0$. If the numerator approaches a finite limit as $\omega \to 0$, then the real part of $\sigma_{xx}(\omega)$ will contain a delta function contribution $D_s\delta(\omega)$ with Drude weight
\begin{equation}
    \frac{D_s}{\pi e^2} = \frac{\rho_s^*}{4\pi} = -\langle K_x\rangle + \Pi_{xx}(q_x=0, q_y \to 0, \omega=0), \label{Ds}
\end{equation}
where $\rho^*_s$ is the superfluid density. Crucially, we first set $\omega =0$ and $q_x = 0$. \emph{Then} we take the limit $q_y \to 0$.
The kinetic energy density is straightforward to compute:
\begin{equation}
    \begin{aligned}
    \langle K_x\rangle &= -\frac{t}{N}\sum_{i,\sigma}\sum_{n(E_n\geq0)}\left\{f(E_n)\left[u^{n*}_{i+\hat{x},\sigma}u_{i,\sigma}^n + \text{c.c} \right] + (1-f(E_n))\left[v^{n}_{i+\hat{x},\sigma}v_{i,\sigma}^{n*} + \text{c.c} \right]\right\}\\
    &= -\frac{2t}{N}\sum_{i}\sum_{n(E_n\geq0)}\left\{f(E_n)\left[u^{n*}_{i+\hat{x},\sigma}u_{i,\sigma}^n + \text{c.c} \right] + (1-f(E_n))\left[v^{n}_{i+\hat{x},\sigma}v_{i,\sigma}^{n*} + \text{c.c} \right]\right\}
\end{aligned}
\end{equation}
In going to the second line, we account for the sum over spins with a factor of 2. Next, let us compute the current-current term. We aim to calculate 
\begin{equation}
    \begin{aligned}
    [J_x^P(\bq,t),J_x^P(-\bq,0)] &= [J_{x,\uparrow}^P(\bq,t)+J_{x,\downarrow}^P(\bq,t),J_{x,\uparrow}^P(-\bq,0)+J_{x,\downarrow}^P(-\bq,0)]\\
    &= [J_{x,\uparrow}^P(\bq,t),J_{x,\uparrow}^P(-\bq,0)] + [J_{x,\uparrow}^P(\bq,t),J_{x,\downarrow}^P(-\bq,0)]\\
    &~~~~~+[J_{x,\downarrow}^P(\bq,t),J_{x,\uparrow}^P(-\bq,0)]+[J_{x,\downarrow}^P(\bq,t),J_{x,\downarrow}^P(-\bq,0)]\\
    &= [J_{x,\uparrow}^P(\bq,t),J_{x,\uparrow}^P(-\bq,0)] +[J_{x,\downarrow}^P(\bq,t),J_{x,\downarrow}^P(-\bq,0)]
\end{aligned}
\end{equation}
In going to the last line, we use the fermion commutation relation. Because our system has spin-flip symmetry, we can rewrite this as
\begin{equation}
    [J_x^P(\bq,t),J_x^P(-\bq,0)] = 2[J_{x,\sigma}^P(\bq,t),J_{x,\sigma}^P(-\bq,0)],
\end{equation}
where $\sigma$ can represent either $\uparrow$ or $\downarrow$; for our purposes, it does not matter. We can rewrite the current-current correlator as
\begin{equation}
[J_{x,\sigma}^P(\bq,t),J_{x,\sigma}^P(-\bq,0)] = (it)^2\sum_{i}\sum_{j}e^{-i\bq(\br_i-\br_j)}[\hat{A}+\hat{B}+\hat{C}+\hat{D}]
\end{equation}
where we've defined
\begin{equation}
    \begin{aligned}
    \hat{A} &= [c^\dagger_{i+\hat{x},\sigma}(t)c_{i,\sigma}(t),c^\dagger_{j+\hat{x},\sigma}(0)c_{j,\sigma}(0)]\\
    \hat{B} &= -[c^\dagger_{i+\hat{x},\sigma}(t)c_{i,\sigma}(t),c^\dagger_{j,\sigma}(0)c_{j+\hat{x},\sigma}(0)]\\
    \hat{C} &= -[c^\dagger_{i,\sigma}(t)c_{i+\hat{x},\sigma}(t),c^\dagger_{j+\hat{x},\sigma}(0)c_{j,\sigma}(0)]\\
    \hat{D} &= [c^\dagger_{i,\sigma}(t)c_{i+\hat{x},\sigma}(t),c^\dagger_{j,\sigma}(0)c_{j+\hat{x},\sigma}(0)]\\
\end{aligned}
\end{equation}
Upon rewriting these operators in the BdG basis, we end up with an expression for the current-current correlator in momentum-frequency space
\begin{equation}
    \Pi_{xx}(\bq) = \frac{1}{N}\sum_{n,m}\left\{\frac{A_{nm}(\bq)\left[A^*_{nm}(\bq)+D_{nm}(-\bq)\right]}{E_n-E_m} \right\} \times \left[f(E_n)-f(E_m) \right]
\end{equation}
Here, the sum is over \textit{all} energy eigenvalues. To find the superfluid stiffness, we must then take the limit $q_x=0, q_y\to 0$, as described in Equation \ref{Ds}. The transition temperature is directly proportional to this quantity:
\begin{equation}
    T_c = \frac{D_s}{\pi e^2} = -\langle K_x\rangle + \Pi_{xx}(q_x=0, q_y\to 0, \omega=0)
\end{equation}
\section{Uniform susceptibility} \label{appendix:uniform_sus}
To understand how the uniform susceptibility is relevant to understanding weak coupling instabilities, let us write the linearized gap equation in the following form,
\begin{align}
\lambda F_{\br,\br+\bdelta} = - \sum_{\br' , \bdelta'} V_{\br,\br+\bdelta} \chi_{\br,\br+\bdelta;\br' ,\br'+\bdelta'} F_{\br',\br'+\bdelta'}.
\end{align}
where $F_{\br,\br+\bdelta}$ is a normalized eigenvector with eigenvalue $\lambda$, and $\chi$ is the susceptibility in either the s- or p-wave channel. If there is a weak coupling instability, then $\lambda \rightarrow +\infty$ as the temperature $T\rightarrow 0$. To proceed let us expand $F_{\br,\br+\bdelta}$ according to the relative coordinate of the function, $\bdelta$, and $\bm{r}$, which can be treated as the center of mass coordinate. Fourier transforming the gap function with respect to $\bm{r}$, we have that
\begin{equation}
    F_{\br,\br+\bdelta} = \frac{1}{L^{d/2}} \sum_{\bm{k}} F_{\bm{k},\bdelta} e^{i \bm{k}\cdot \br}
\end{equation}
In terms of the Fourier components, the linearized gap equation is
\begin{align}
\lambda F_{\bm{k},\bdelta} &= - \sum_{ \bm{k}'} \sum_{\br, \br' , \bdelta'} V_{\br,\br+\bdelta} \chi_{\br,\br+\bdelta;\br' ,\br'+\bdelta'} F_{\bm{k},\bdelta'}e^{i \bm{k}\cdot \br'}e^{i (\bm{k}'\cdot \br' - \bm{k}\cdot \br)}
\end{align}
The $\bm{k} = 0$ component of the solution can then be written in the suggestive form as
\begin{align}
\lambda F_{0,\bdelta} &= - \sum_{\br, \br' , \bdelta'} V_{\br,\br+\bdelta} \chi_{\br,\br+\bdelta;\br' ,\br'+\bdelta'} F_{0,\bdelta'}e^{i \bm{k}\cdot \br'}  -  \sum_{ \bm{k}'\neq 0} \sum_{\br, \br' , \bdelta'} V_{\br,\br+\bdelta} \chi_{\br,\br+\bdelta;\br' ,\br'+\bdelta'} F_{\bm{k},\bdelta'}e^{i \bm{k}\cdot \br'}.
\label{eq:kzeroLGE}\end{align}
From this we find that if the uniform susceptibility, 
\begin{align}
    \chi^0_{\bdelta,\bdelta'} &= \sum_{\br\br'}\chi_{\br,\br+\bdelta;\br',\br'+\bdelta'}
\end{align}
diverges as $T\rightarrow 0$, then $\lambda$ must also diverge in this limit, indicating a weak coupling instability. 

To reach this conclusion, we have made three notable assumptions. First, that the interaction $V_{\br,\br+\bdelta}$ is non-zero and only depends on the relative coordinate $\bdelta$. This is simply a choice of interactions. Second, there is not another divergence in the sum over $\bm{k}\neq 0$ that exactly cancels out the divergence of the uniform susceptibility. This assumption innocuous, as an exact cancellation of two divergent terms would be fine-tuned. Third, that $F_{0,\delta}\neq 0$. It should be noted that there are two important, and physically relevant, cases where $F_{0,\delta} = 0$. First, it can be the case that the superconductivity oscillates with a fixed wave-vector, i.e., it is a pair density wave. In this case, $F_{0,\delta} = 0$ and $|F_{\bm{k},\delta}|$ is an $\mathcal{O}(1)$ number for some finite number of $\bm{k}$'s and is zero otherwise. Second, it can be the case that the superconductivity is real space localized. For example, where, in real space, $F_{\br,\br+\bdelta} = \delta_{\br,\bm{R}}$ for some fixed $\bm{R}$. the Fourier transform of this is then $F_{\bm{k},\delta} = \frac{1}{L^{d/2}} e^{i\bm{R}\cdot \bm{k}} $ in $d$-dimensions. Therefore, in the thermodynamic limit, $L \rightarrow \infty$, $F_{\bm{k},\bm{\delta}}\rightarrow 0$. Hence, a weak coupling instability towards either a pair density wave or localized state will not lead to a divergence in the uniform susceptibility.

\subsection*{Averaging procedure}
In the main text, we are compute the following quantity:
\begin{equation}
   \begin{aligned}
\Bar{{\chi}}^0_{\bdelta,\bdelta'} = \frac{1}{(NL)^{6}}\sum_{{{m}}{{n}}}&\frac{1-f(E_{{n}})-f(E_{{m}})}{E_{{n}}+E_{{m}}}\sum_{s=1}^N\sum_{\bk\bp} e^{i\bdelta\cdot(\bk+\bphi_s)}{U}^{*{{n}}}_{-\bk}(-\bphi_s,\bm{\theta}_s){U}^{*{{m}}}_{\bk}(\bphi_s,\bm{\theta}_s)\\
       &\times\Big[e^{-i\bdelta\cdot(\bp+\bphi_s)}{U}^{{n}}_{-\bp}(-\bphi_s,\bm{\theta}_s){U}^{{m}}_{\bp}(\bphi_s,\bm{\theta}_s)\pm e^{-i\bdelta\cdot(\bp-\bphi_s)}{U}^{{n}}_{\bp}(\bphi_s,\bm{\theta}_s){U}^{{m}}_{-\bp}(-\bphi_s,\bm{\theta}_s)\Big].
   \end{aligned}
\end{equation}
Here, we use ${H}(\bphi_s,\bm{\theta}_s)$ to refer to the quasiperiodic Hamiltonian defined on a lattice of size ${L}$ where the hopping terms have a phase factor $\bphi_s$ and quasiperiodic potential has an offset $\bm{\theta}_s$. We use $N$ to denote the number of realizations of ${L}$  and $\bphi_s$. The ${n}$th eigenstate is denoted ${U}^{{n}}_{\bk}(\bphi_s,\bm{\theta}_s)$. The sum over $s$ corresponds to averaging over twists and potential offsets. For simplicity, we set $\bm{\theta}_s = \bphi_s$, and average over both twists and potential offsets simultaneously. Specifically, we average over $N\approx 35$ realizations of the Hamiltonian at every $v$, for both the quasiperiodic and randomly-disordered lattices. In the case of quasiperiodicity, we generate potential realizations by randomly sampling a single $\bphi$ for each realization of the lattice. In the randomly-disordered case, we sample $\bphi$ randomly on each site. We subsequently average over all ${\chi}^0_{\bdelta,\bdelta'}$ at fixed $v$. We also compute the standard error of the mean, which appears as a ribbon in Figure \ref{fig:uniform_sus}.

We will now explain how averaging over different values of $\bphi_s$ allows us to approximate the behaviour of much larger systems. Here, we will assume that the potential offset $\bm{\theta} $ vanishes, as the value of $\bm{\theta} $ does not affect this discussion. For a system of size $\tilde{L}$ with untwisted boundardy conditions, the uniform susceptibility is
\begin{equation}
    \begin{aligned}
\tilde{\chi}^0_{\bdelta,\bdelta'} &= \frac{1}{\tilde{L}^{6}}\sum_{\br\br'}\chi_{\br,\br+\bdelta;\br',\br'+\bdelta'}\\
    &= \frac{1}{\tilde{L}^{6}}\sum_{\br\br'}
    \sum_{\tilde{m}\tilde{n}}
    \frac{1-f(E_{\tilde{n}})-f(E_{\tilde{m}})}{E_{\tilde{n}}+E_{\tilde{m}}}
    \Big[\tilde{U}^{*\tilde{n}}_{\br} \tilde{U}^{\tilde{n}}_{\br'} 
    \tilde{U}^{*\tilde{m}}_{\br+\bdelta}\tilde{U}^{\tilde{m}}_{\br'+\bdelta'} \pm \tilde{U}^{*\tilde{n}}_{\br}\tilde{U}^{\tilde{n}}_{\br'+\bdelta'}\tilde{U}^{*\tilde{m}}_{\br+\bdelta}\tilde{U}^{\tilde{m}}_{\br'}\Big].\label{chi_realspace}
\end{aligned}
\end{equation}
We will now argue for approximating this quantity using a system of size ${L}$, such that $N{L} = \tilde{L}$ for some large integer $N$. The Hamiltonian for the full system of size $\tilde{L}$ will be given by $\tilde{H}$, while the Hamiltonian for the size-${L}$ system will be given by ${H}$. The kinetic parts of both $\tilde{H}$ and ${H}$ are the same, while the quasiperiodic parts differ on account of the rational approximant used to generate the potential. Let us use $\mathbf{B}_{\tilde{L}}$ to denote the rational approximant to the quasiperiodic ordering vector on a system of size $\tilde{L}$. If the rational approximants are both close to the actual irrational ordering vector, then $\mathbf{B}_{\tilde{L}} \approx \mathbf{B}_{{L}}$. 

Since $\tilde{H}$ corresponds to a $\tilde{L}^3$ matrix, computational limitations prevent us from calculating the uniform susceptibility of $\tilde{H}$ when $\tilde{L}$ is large. However, we shall argue that the uniform susceptibility of $\tilde{H}$ can be well approximated by considering $\bphi_s$ averages of ${H}(\bphi_s)$, the quasiperiodic Hamiltonian defined on the system of size ${L}$, where we have attached a phase of $\phi_{s,i}$ to the hopping term in the $\hat{u}_i$ direction. Under a gauge transformation, this is equivalent to twisting the boundary condition in the $\hat{u}_i$ direction by ${L}\phi_{s,i}$. Here, we let $\bphi$ take on the following values:
\begin{equation}
    \phi_{s,i} = \frac{2\pi s_i}{L} = \frac{2\pi s_i}{N {L}} ~~~~~ s_i\in\{1,...,N\}
\end{equation}
Specifically, we are calculating the Fourier transform of the uniform susceptibility of ${H}$, averaged over phases $\bphi$:
\begin{equation}
    \begin{aligned}
\Bar{{\chi}}^0_{\bdelta,\bdelta'} 
    = \frac{1}{L^{6}}&\sum_{mn}\frac{1-f(E_{{n}})-f(E_{{m}})}{E_{{n}}+E_{{m}}}\\
    &~~~~~~~~~~\times\sum_{s=0}^{N-1}\sum_{\br\br'=0}^{{L}-1} \Big[{U}^{*{{n}}}_{\br}(\bphi_s) {U}^{{n}}_{\br'}(\bphi_s) {U}^{*{{m}}}_{\br+\bdelta}(\bphi_s){U}^{{m}}_{\br'+\bdelta'}(\bphi_s) \pm {U}^{*{{n}}}_{\br}U^{{{n}}}_{\br'+\bdelta'}(\bphi_s){U}^{*{{m}}}_{\br+\bdelta}(\bphi_s){U}^{{m}}_{\br'}(\bphi_s)\Big]\label{chi_realspace}
\end{aligned}
\end{equation}
where we use ${U}(\bphi_s)$ are the eigenstates of ${H}(\bphi_s))$. Since ${H}$ is a ${L}^3 \ll \tilde{L}^3$ matrix, this leads to a significant computational improvement. We will argue below that $\Bar{{\chi}}^0_{\bdelta,\bdelta'} $ is a good approximation to $\tilde{\chi}^0_{\bdelta,\bdelta'}$. In practice, we replace the sum over discrete $\bm{\phi}_s$ with random sampling, to avoid any unphysical periodic structures. 

\subsubsection{Ballistic phase}
We will first show that $\tilde{\chi}^0_{\bdelta,\bdelta'}$ is well approximated by $\tilde{\chi}^0_{\bdelta,\bdelta'}$ in the ballstic phase. We begin by considering the uniform susceptibility in momentum space. Here, we consider $\bdelta,\bdelta'=0$ for simplicity, though these results can be easily generalized to nonzero $\bdelta,\bdelta'$. Because the eigenstates are localized in momentum space in the ballistic phase, the dominant contributions come from terms with $\bk\approx\bp$, such that
\begin{align}
\tilde{\chi}^0_{0,0} 
    &=\frac{1}{\tilde{L}^{6}} \sum_{\bk}\sum_{\tilde{m}\tilde{n}}\frac{1-f(E_{\tilde{n}})-f(E_{\tilde{m}})}{E_{\tilde{n}}+E_{\tilde{m}}}\tilde{U}^{*\tilde{n}}_{-\bk}
    \tilde{U}^{*\tilde{m}}_{\bk}\Big[\tilde{U}^{\tilde{n}}_{-\bk}\tilde{U}^{\tilde{m}}_{\bk}\pm \tilde{U}^{\tilde{n}}_{\bk}\tilde{U}^{\tilde{m}}_{-\bk}\Big]\label{unifsus_fullL}
\end{align}
Separately, consider the system of size $N^3{L}^3 = \tilde{L}^3$ composed of stitching together $N^3$ copies of the Hamiltonian $H$. We will refer to this Hamiltonian as $H'$.  $H'$ is almost equivalent to $\tilde{H}$, except for the fact that the potential is constructed using rational approximant $\mathbf{B}_{{L}}$ as opposed to $\mathbf{B}_{\tilde{L}}$. Since $\mathbf{B}_{{L}}\approx \mathbf{B}_{\tilde{L}}$, it follows that in the ballsitic phase (where the energy eigenstates are localized in momnetum space) the eigenstates $U'$ of $H'$ are approximately equal to the eigenstates $\tilde{U}$ of the fully quasiperiodic system $\tilde{H}$. This implies that we can approximate $\tilde{\chi}^0_{\bdelta,\bdelta'}$ with the susceptibility of $H'$. Since $H'$ is $N^3$ copies ${H}$, the eigenstates of $H'$ can be written in terms of the eigenvectors of ${H}$ summing over twisted boundary conditions. The susceptibility of $H'$ is therefore equal to $\Bar{{\chi}}^0_{\bdelta,\bdelta'}$. 

\subsubsection{Diffusive \& localized regimens}
In the localized or diffusive regimes (i.e., $v > v_{c1}$), there is a characteristic length scale $l$ such that the real space susceptibility calculated using $\tilde{H}$, $\tilde{\chi}_{\br,\br+\bdelta;\br',\br'+\bdelta'}$ falls off exponentially for $|\br - \br'| > l$ \cite{john}. In the localized regime, $l$ is the real space localization length, in the diffusive regime, $l$ is thermal coherence length $\sqrt{\frac{\hbar D}{T}}$, which is finite at finite temperature. Hence, if ${L}\gg l$, then the uniform susceptibility calculated using $\tilde{H}$ and ${H}(\bm{\phi} = 0)$ should be equivalent. Furthermore, if  ${L}\gg l$ twisting boundary conditions will not affect the susceptibility, as twisting boundary conditions will only affect the susceptibility when $\br$ and $\br'$ are far separated. Based on these two observations, we conclude that $\tilde{\chi}^0_{\bdelta,\bdelta'}$ will be well approximated by $\Bar{{\chi}}^0_{\bdelta,\bdelta'}$. 

\begin{figure}
\centering
\includegraphics[width=0.95\textwidth]{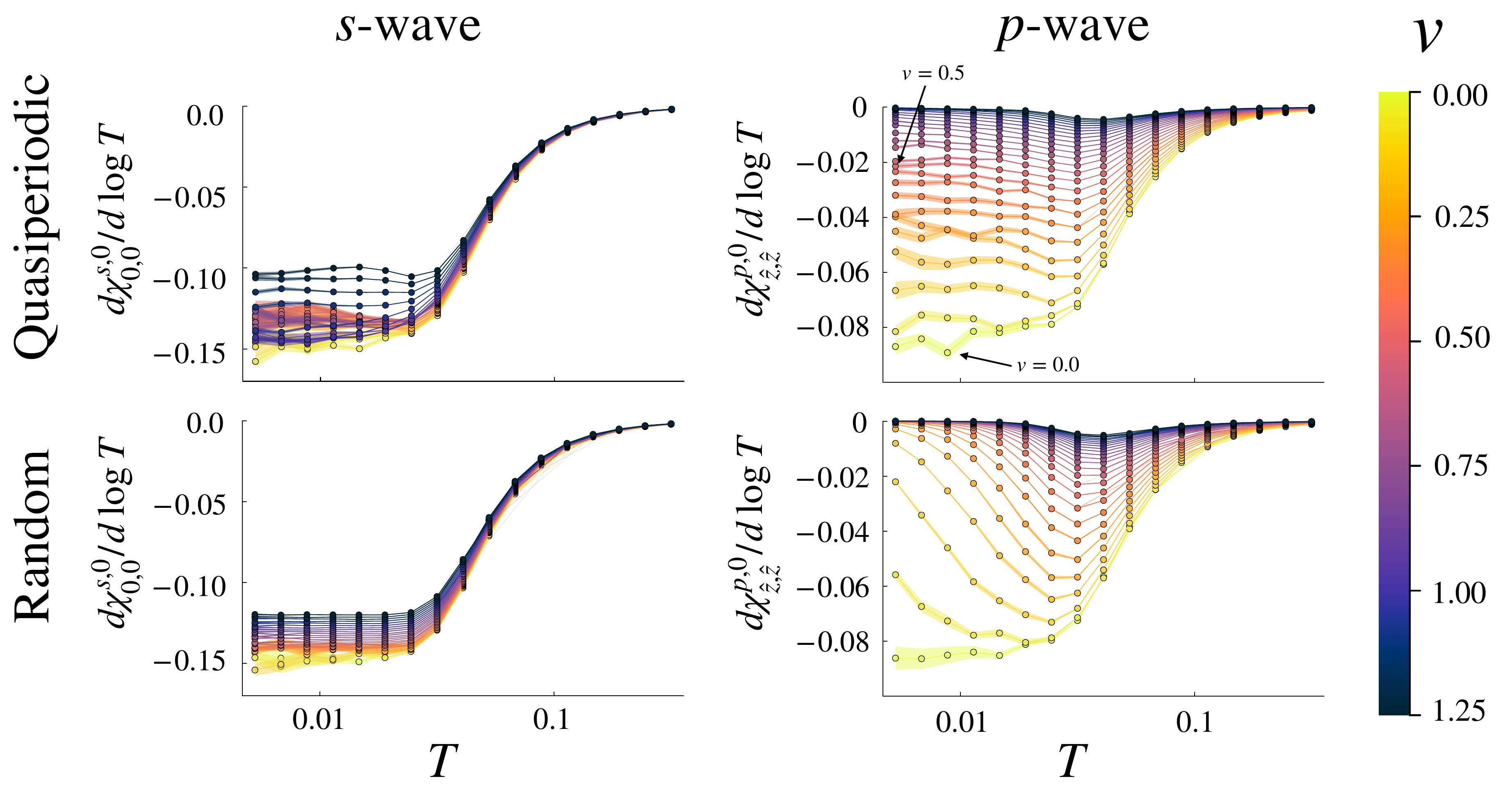}
\caption{$d\chi^0_{\bdelta,\bdelta'}/d\log T$ calculated as a function of temperature for a range of disorder strengths for both $s$- (left column) and $p$-wave (right column) superconductivity. These data are shown for both quasiperiodic (top row) and random (bottom row) disorder, and are computed on a $23\times 23\times 23$ lattice with $\theta=\pi/7$, $Q=(\sqrt{2}-1)/2$, and $0.75=\mu/(1+v)$. Results are averaged over $N\approx 35$ potential realizations with error ribbons showing the standard error of the mean.}
\label{fig:uniform_sus_appendix}
\end{figure}
\subsection*{Numerics}
In Figure \ref{fig:uniform_sus_appendix}, we also plot the $d\chi^0_{\bdelta,\bdelta'}/d\log T$ data as a function of temperature $T$ rather than potential strength $v$. We show that the disorder-free case ($v=0$, yellow curve) saturates to a constant value for both the quasiperiodic and random disorder potentials. Once we introduce a finite amount of quasiperiodicity or disorder, however, we start to see differences between these two cases. Any amount of random disorder will start to suppress the logarithmic divergence of the unconventional superconductor, as captured by the increase in the derivative of the uniform susceptibility as the temperature is lowered. 

\end{document}